\documentclass[aps,prb,showpacs,preprintnumbers,amsmath,amssymb,twocolumn,letterpaper,superscriptaddress, 10pt]{revtex4-1}
\usepackage{amssymb}
\usepackage{ifpdf}
\usepackage[utf8]{inputenc}
\usepackage{graphicx}
\usepackage{times}
\usepackage{dcolumn}
\usepackage{textcomp}
\usepackage{amsmath}
\usepackage{array}
\usepackage{color}
\usepackage{bibunits}

\makeatletter
\newcommand*{\balancecolsandclearpage}{%
  \close@column@grid
  \clearpage
  \twocolumngrid
}
\makeatother

\newcommand{\ug}{UGe$_{2}$}
\newcommand{\mB}{$\mu_{\rm B}$}
\newcommand{\tc}{$T_{\rm c}$}
\newcommand{\tx}{$T_{\rm x}$}
\newcommand{\pc}{$p_{\rm c}$}
\newcommand{\px}{$p_{\rm x}$}
\newcommand{\hw}{$\hbar\omega$}

\newcommand{\vQ}{$\mathbf{q}$}

\begin{document}

\title{Ultra-High Resolution Neutron Spectroscopy of Low-Energy Spin Dynamics in \ug}

\author{F. Haslbeck}
\affiliation{Physik-Department, Technische Universit\"at M\"unchen, D-85748 Garching, Germany}
\affiliation{Institute for Advanced Study,
Technische Universit\"at M\"unchen, D-85748 Garching, Germany}

\author{S. S\"aubert}
\affiliation{Physik-Department, Technische Universit\"at M\"unchen, D-85748 Garching, Germany}
\affiliation{Heinz Maier-Leibnitz Zentrum Garching, Technische Universit\"at M\"unchen, D-85748 Garching, Germany}

\author{M. Seifert}
\affiliation{Physik-Department, Technische Universit\"at M\"unchen, D-85748 Garching, Germany}
\affiliation{Heinz Maier-Leibnitz Zentrum Garching, Technische Universit\"at M\"unchen, D-85748 Garching, Germany}

\author{C. Franz}
\affiliation{Heinz Maier-Leibnitz Zentrum Garching, Technische Universit\"at M\"unchen, D-85748 Garching, Germany}

\author{M. Schulz}
\affiliation{Heinz Maier-Leibnitz Zentrum Garching, Technische Universit\"at M\"unchen, D-85748 Garching, Germany}

\author{A. Heinemann}
\affiliation{German Engineering Materials Science Centre (GEMS) at Heinz Maier-Leibnitz Zentrum (MLZ),
Helmholtz-Zentrum Geesthacht GmbH, D-85747 Garching, Germany}

\author{T. Keller}
\affiliation{Max-Planck-Institut f\"{u}r Festk\"{o}rperforschung, D-70569 Stuttgart, Germany}
\affiliation{Max Planck Society Outstation at the Forschungsneutronenquelle Heinz Maier-Leibnitz (MLZ), D-85747 Garching, Germany}

\author{Pinaki Das}
\altaffiliation[Current address: ]{Division of Materials Sciences and Engineering, Ames Laboratory, U.S. DOE, Iowa State University, Ames, Iowa 50011, USA}
\affiliation{Los Alamos National Laboratory, Los Alamos, New Mexico 87545, USA}

\author{J. D. Thompson}
\affiliation{Los Alamos National Laboratory, Los Alamos, New Mexico 87545, USA}

\author{E. D. Bauer}
\affiliation{Los Alamos National Laboratory, Los Alamos, New Mexico 87545, USA}

\author{C. Pfleiderer}
\affiliation{Physik-Department, Technische Universit\"at M\"unchen, D-85748 Garching, Germany}

\author{M. Janoschek}
\email[Corresponding Author: ]{marc.janoschek@psi.ch}
\affiliation{Institute for Advanced Study, Technische Universit\"at M\"unchen, D-85748 Garching, Germany}
\affiliation{Los Alamos National Laboratory, Los Alamos, New Mexico 87545, USA}
\affiliation{Laboratory for Scientific Developments and Novel Materials, Paul Scherrer Institut, Villigen PSI, Switzerland}

\date{\today}

\begin{abstract}
    Studying the prototypical ferromagnetic superconductor \ug\ we demonstrate the potential of the Modulated IntEnsity by Zero Effort (MIEZE) technique---a novel neutron spectroscopy method with ultra-high energy resolution of at least 1~$\mu$eV---for the study of quantum matter. We reveal purely longitudinal spin fluctuations in \ug\ with a dual nature arising from $5f$ electrons that are hybridized with the conduction electrons. Local spin fluctuations are perfectly described by the Ising universality class in three dimensions, whereas itinerant spin fluctuations occur over length scales comparable to the superconducting coherence length, showing that MIEZE is able to spectroscopically disentangle the complex low-energy behavior  characteristic  of  quantum  materials.  
\end{abstract}

\vskip2pc
 
\maketitle
\begin{bibunit}
\section{Introduction}

Ultra-slow spin dynamics represent a key characteristic of quantum matter such as quantum spin liquids~\cite{Han:2012}, electronic nematic phases~\cite{Fernandes:2014}, topological spin textures~\cite{Muehlbauer:2009, Janoschek:2013}, non-Fermi liquid behavior~\cite{Aaronson:1995, Schroeder:2000, Kadowaki:2006, Knafo:2009}, and unconventional superconductivity~\cite{Scalapino:2012}. For the clarification of these phenomena spectroscopic methods with excellent momentum and energy resolution are required, as key characteristics emerge typically in the low milli-Kelvin range. In principle, neutron scattering is ideally suited for studies of the relevant spin excitations. However, the typical energy resolution of conventional neutron spectroscopy corresponds to Kelvin temperatures. Although techniques such as neutron spin-echo spectroscopy offer ultra-high resolution of sub-$\mu{\rm eV}$, they are incompatible with conditions that depolarize neutron beams such as ferromagnetism (FM), superconductivity or large magnetic fields. 

The discovery of superconductivity in the FM state of \ug\ highlights the combination of scientific and experimental challenges that arise in the study of the complex low-energy behavior of quantum matter that characteristically emerges due to the competition of high-energy atomic energy scales ~\cite{Tsvelik:2003}. Namely, actinide-based compounds such as \ug\ are formidable model systems, where the hybridization of itinerant $d$ and localized $f$ electrons drives low-energy excitations that mediate a multitude of novel states~\cite{Dagotto:2005, Pfleiderer:2009, Moore:2009}. Traditionally the concomitant subtle reconstruction of the electronic structure has been studied via the charge channel, which fails to provide the required high resolution~\cite{Im:2008,Schmidt:2010}. Exploiting in contrast the spin channel recent advances in neutron spectroscopy provided new insights \cite{Butch:2015, Janoschek:2015, Goremychkin:2018}.

Using an implementation of neutron resonance spin echo (NRSE) spectroscopy that is insensitive to depolarizing conditions, namely the so-called Modulated IntEnsity by Zero Effort (MIEZE) \cite{Gaehler:1992}, we report a study of \ug\ in which we identify the enigmatic low-energy excitations as an unusual combination of fluctuations attributed normally either to itinerant or localized electrons in an energy and momentum range comparable to the superconducting coherence length and ordering temperature. As the superconductivity in {\ug} represents a prototypical form of quantum matter, our study underscores also the great potential of the MIEZE technique in studies of quantum matter on a more general note.

At ambient pressure {\ug} displays ferromagnetism with a large Curie temperature, \tc = 53 K, and a large ordered moment, $\mu_{\textrm{FM1}}=$1.2 \mB (FM1) ~\cite{Saxena:2001, Pfleiderer:2002}. Under increasing pressure FM order destabilizes, accompanied by the emergence of a second FM phase below \tx $<$ \tc\, where $\mu_{\textrm{FM2}}=$1.5 \mB\ (FM2). The FM2 and FM1 phases vanish discontinuously at \px\ $\approx$~12.2 kbar and  \pc\ $\approx$~15.8 kbar, respectively~\cite{Pfleiderer:2002}, while superconductivity emerges between $\approx$~9 kbar $<$~\px\ and {\pc}. Evidence for a microscopic coexistence of FM order and superconductivity makes \ug\ a candidate for p-wave pairing, where the Cooper pairs form spin triplets~\cite{Saxena:2001}. This p-wave superconductivity is believed to be mediated by an abundance of low-lying longitudinal spin fluctuations associated with a FM quantum phase transition (QPT), where transverse spin fluctuations are theoretically known to break spin-triplet pairing~\cite{Fay:1980}. Neutron triple-axis spectroscopy (TAS) at ambient pressure indeed identified predominantly longitudinal spin fluctuations in \ug~\cite{Huxley:2003}, but failed to provide insights into the character of the fluctuations in the momentum and energy range comparable to the superconducting coherence length and transition temperature, respectively.

\begin{figure}[th]
	\begin{center}
		\includegraphics[width=0.9\columnwidth]{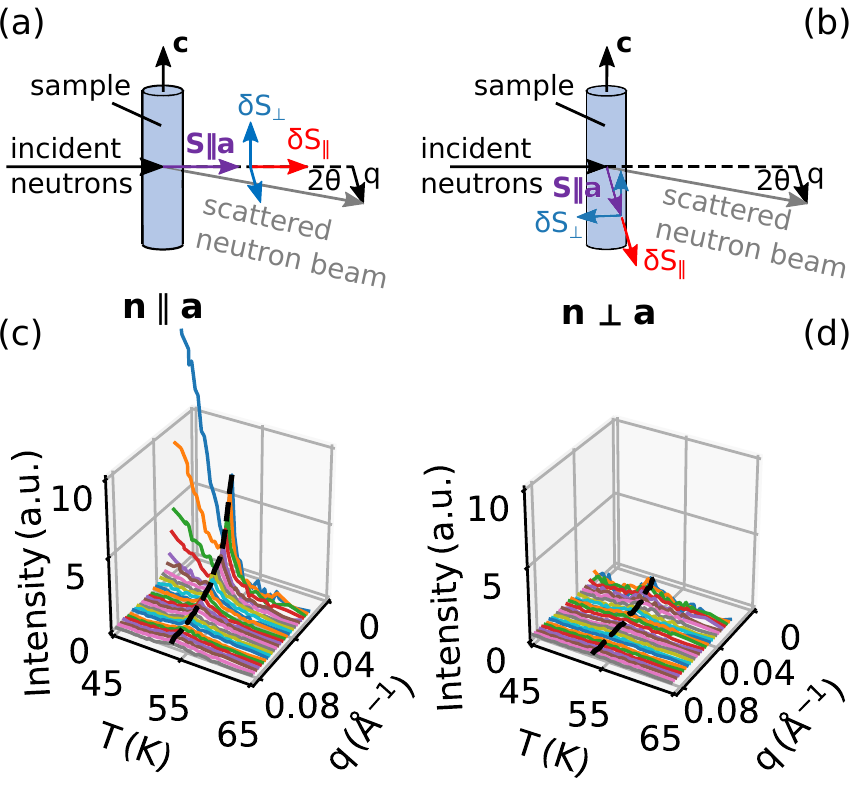}
		\caption{Magnetic intensity in \ug\ near the Curie temperature $T_\textrm{C}=52.7$~K. (a) and (b) show two experimental configurations with the $a$ axis parallel or perpendicular to the incident neutron beam $\mathbf{n}$, respectively, used to differentiate longitudinal from transverse spin fluctuations (see text). (c) and (d) show the observed energy-integrated intensities for $\mathbf{n}\parallel a$ and $\mathbf{n}\perp a$ as a function of temperature $T$ and momentum transfer \vQ. The black dashed line marks $T_\textrm{C}$.}
		\label{fig:setup-and-energy-integrated}
	\end{center}
\end{figure}

Prior to our study the interplay of seemingly conflicting ingredients of the spectrum of spin fluctuations were unresolved. On the one hand, the strong Ising anisotropy promotes longitudinal spin fluctuations as typically attributed to localized electrons in the presence of strong to spin-orbit coupling. This is contrasted, on the other hand, by the notion of Cooper pairs and a well developed, strongly exchange-split Fermi surface  ~\cite{Fay:1980, Monthoux:1999}. Consistent with this dichotomy characteristic of p-wave superconductivity, our ultra-high resolution  data reveals that the low-energy spin fluctuations of \ug\ reflect a subtle interplay of itinerant and local electronic degrees of freedom on scales comparable to the superconductivity.

\section{Experimental Methods}

NRSE achieves extreme energy resolution by encoding the energy transfer \hw\ of the neutrons in their polarization as opposed to a change of wavelength. However, FM domains, Meissner flux expulsion or applied magnetic fields typically depolarize the beam. We used therefore a novel NRSE technique, so called MIEZE, implemented at the instrument RESEDA at the Heinz Maier-Leibnitz Zentrum (MLZ) \cite{HAUSSLER:2007, Kindervater:2015, Krautloher:2016}. Generating an intensity modulated beam by means of resonant spin flippers and a spin analyzer in front of the sample the amplitude of the intensity modulation assumes the role of the NRSE polarization. Because all spin manipulations are performed before the sample, beam depolarizing effects are no longer important. Using incident neutrons with a wavelength $\lambda=6$~\AA~ and $\Delta\lambda/\lambda\approx10$\% provided by a velocity selector, we achieved an energy resolution of $\Delta E\approx 1\,\mu\text{eV}$. MIEZE in small angle neutron scattering (SANS) configuration also provides high momentum \vQ\ resolution of approximately~$0.015$~\AA$^{-1}$. The MIEZE setup is described in the supplemental material~\cite{supplement}.

\begin{figure}[th]
    \begin{center}
        \includegraphics[width=0.9\columnwidth]{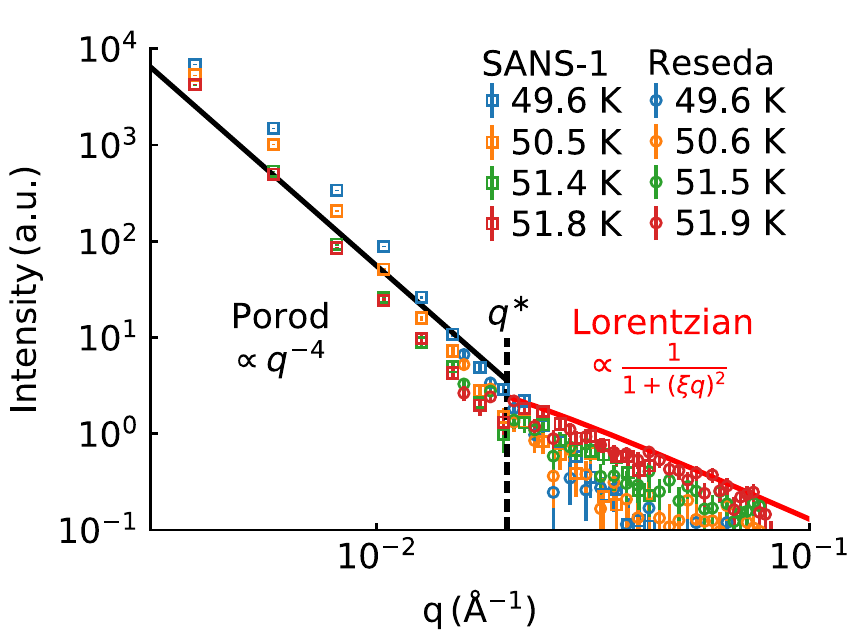}
        \caption{\vQ-dependence of the intensity for selected $T$ below $T_\textrm{C}$ for $\mathbf{n}\parallel a$. Below $q^{\ast}\approx0.02$~\AA$^{-1}$ the intensity is well described by Porod scattering due to ferromagnetic (FM) domains (black solid line), whereas above $q^{\ast}$ a Lorentzian shape due to critical spin fluctuations is observed (red solid line).}
        \label{fig:porod-and-lorentzian}
    \end{center}
\end{figure}
\begin{figure}[h]
    \begin{center}
        \includegraphics[width=0.9\columnwidth]{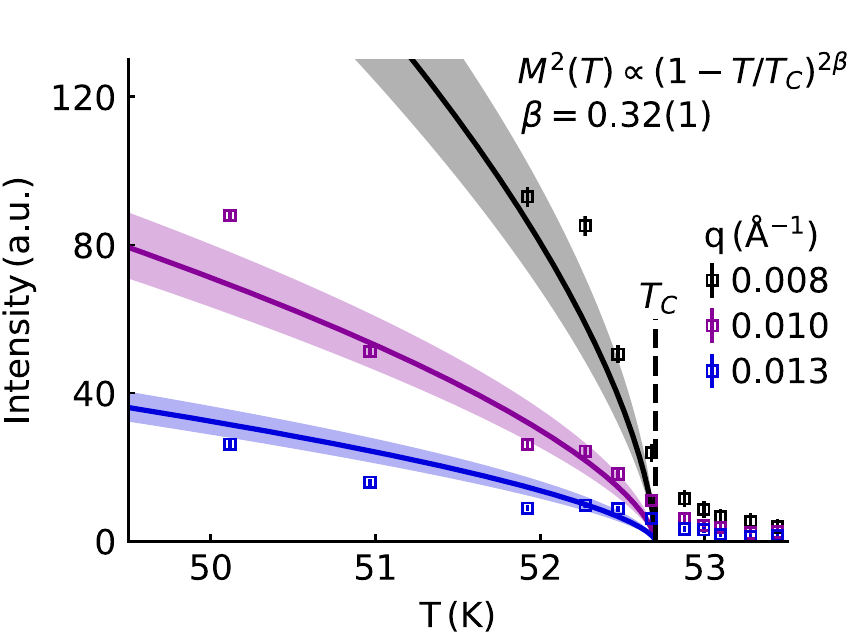}
        \caption{The $T$ dependence of the Porod scattering for $q<q^{\ast}$ follows the FM order parameter $M$ via $M^2(T)\propto(1-\frac{T}{T_\textrm{C}})^{2\beta}$ with $\beta=0.32(1)$ (solid lines). The shaded regions denote the uncertainty of the fit of $\beta$.}
        \label{fig:magnetisation}
    \end{center}
\end{figure}

A high-quality single crystal of \ug\ was grown by the Czochralski technique followed by an annealing similar to Ref. \onlinecite{Huy:2009}. A cylindrical piece with nearly constant diameter of 7 mm and 16 mm length ($m$ = 6 g) with the crystallographic $c$ axis approximately parallel to the cylinder axis was cut for the MIEZE experiments. The sample was oriented using neutron Laue diffraction so that $c$ was perpendicular to the scattering plane. The Laue images also confirm a high-quality single-grain sample~\cite{supplement}. Neutron depolarization imaging measurements ~\cite{supplement} of the same sample reveal that the magnetic properties of the crystal are completely homogeneous with a Curie temperature $T_\textrm{C}$ = 52.68(3) K demonstrating that this sample is optimal for the investigation of critical spin fluctuations. Magnetic susceptibility measurements were performed on a small piece ($m$ = 36 mg) of the same sample in a Quantum Design magnetic property measurement system (MPMS).

\section{Results and Discussion}

The magnetic cross-section is related to the imaginary part of the dynamical magnetic susceptibility $\chi_{ij}''(\mathbf{q}, \omega)$ via
\begin{equation}
	\frac{d^2\sigma}{d\Omega d\omega}\propto\frac{k_f}{k_0}(\delta_{ij}-\hat{q}_{i}\hat{q}_{j})|F_{\mathbf{q}}|^2[n(\omega)+1]\chi_{ij}''(\mathbf{q}, \omega),\label{Eq:cross-section}
\end{equation}
where $k_0$ and $k_f$ are the wave vector of the incident and scattered neutrons, respectively. $\hat{q}$ is a unit vector parallel to the scattering vector \vQ\ and $n(\omega)$ is the Bose function. $F_{\mathbf{q}}$ is the uranium magnetic form factor.

\begin{figure}[t]
    \begin{center}
        \includegraphics[width=0.9\columnwidth]{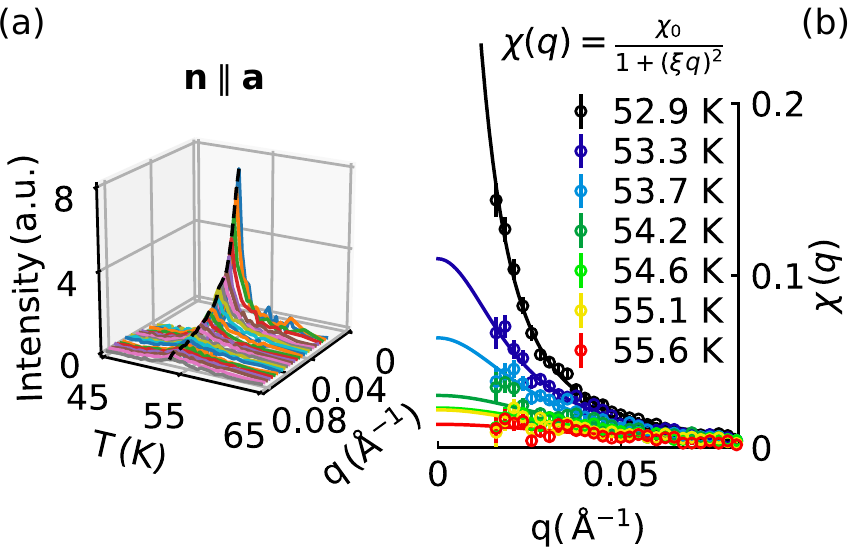}
        \caption{(a) Temperature $T$ and momentum transfer \vQ\ dependence of the critical Ising spin fluctuations in \ug. (b) $q$-dependence of the magnetic susceptibility $\chi(q)$. Solid lines are fits to Eq.~\ref{Eq:q-dependence}.}
        \label{fig:lorentzian}
    \end{center}
\end{figure}

In Fig.~\ref{fig:setup-and-energy-integrated} we show the temperature and \vQ\ dependence of the energy-integrated intensity of the spin fluctuations in \ug\ that was obtained by switching the MIEZE setup off.  Non-magnetic background scattering obtained well above $T_\textrm{C}$ was subtracted from all data sets shown. The temperature scan was carried out with the crystallographic $a$-axis, which is the magnetic easy-axis for \ug\, oriented parallel ($\mathbf{n}\parallel a$) and perpendicular ($\mathbf{n}\perp a$) to the incident neutron beam, respectively. Due to the term $\delta_{ij}-\hat{q}_{i}\hat{q}_{j}$ in Eq.~\ref{Eq:cross-section} neutron scattering is only sensitive to spin fluctuations that are perpendicular to \vQ. Because in SANS configuration \vQ\ is approximately perpendicular to the incident neutron beam, this allows to separate longitudinal ($\delta S_\parallel$) from transverse spin fluctuations ($\delta S_\perp$) as illustrated in Figs.~\ref{fig:setup-and-energy-integrated}(a) and (b). For $\mathbf{n}\parallel a$ both $\delta S_\parallel$ and $\delta S_\perp$ are perpendicular to \vQ. As shown in Fig.~\ref{fig:setup-and-energy-integrated}(c) substantial magnetic intensity is observed for this configuration. In contrast, for $\mathbf{n}\perp a$ only $\delta S_\perp$ is perpendicular to \vQ\, and the vanishingly small signal observed in this case [see ~\ref{fig:setup-and-energy-integrated}(d)] can only come from transverse spin fluctuations. Because of the cylindrical shape of the sample differences in neutron transmission between the two orientations are negligible. As shown in the supplemental material~\cite{supplement}, the small intensity observed for $\mathbf{n}\perp a$ arises from finite \vQ\ resolution, demonstrating that the critical spin fluctuations in \ug\ are solely longitudinal.

\begin{figure}[th]
    \begin{center}
        \includegraphics[width=0.9\columnwidth]{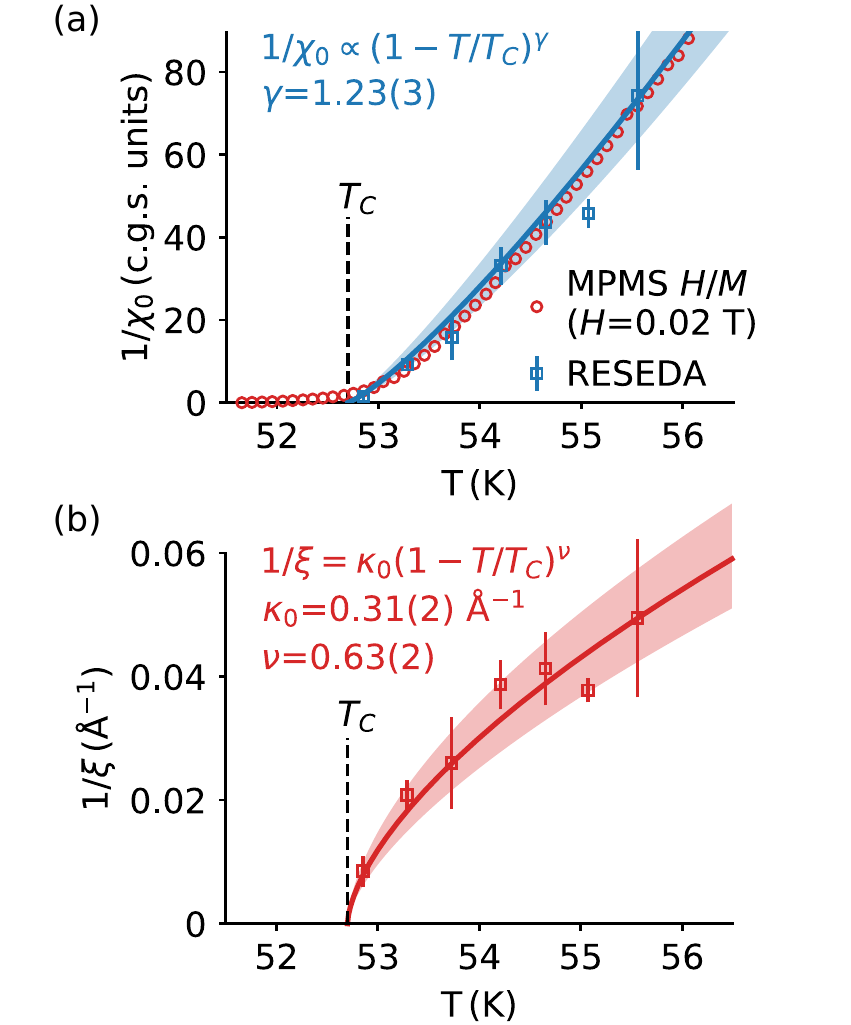}
        \caption{The inverse susceptibility $1/\chi_0$ and inverse correlation length $1/\xi$ as a function of temperature $T$ (near the Curie temperature $T_\textrm{C}=52.7$~K), respectively, resulting from fits in Fig~\ref{fig:lorentzian}. The blue squares in (a) denote the static easy-axis magnetic susceptibility $H/M$ determined with a magnetic field $H$ = 0.1 T. The solid black and red lines are fits to determine the critical exponents for $\chi_0$ and $\xi$ (see text), and the shaded region denotes the uncertainty of the fit.}
        \label{fig:susceptibility-correlation-length}
    \end{center}
\end{figure}

Inspecting the temperature dependence of the integrated intensity for $\mathbf{n}\parallel a$ [see Fig.~\ref{fig:setup-and-energy-integrated}(c)], a pronounced peak is centered at $T_\textrm{C}$ = 52.7 K due to the divergence of critical spin fluctuations. For low $q$ and for $T<T_\textrm{C}$ additional intensity is observed that increases like a magnetic order parameter. Fig.~\ref{fig:porod-and-lorentzian} shows the $q$-dependence of the intensity for a few temperatures below $T_\textrm{C}$. Below $q^{\ast}\approx0.02$~\AA$^{-1}$ the intensity is well-described by a $q^{-4}$ dependence that is characteristic for scattering from FM domains that form below $T_\textrm{C}$~\cite{Lynn:1998, Simon:2002}. To follow this so-called Porod scattering towards lower $q$, we have performed a supporting SANS experiment on the instrument SANS-1 at MLZ (details are described in \cite{supplement}) denoted with square symbols in Fig.~\ref{fig:porod-and-lorentzian}. Observation of Porod scattering down to $q_{\textrm{min}}=0.004$~\AA$^{-1}$ implies the onset of long-range order over length scales $\gg 2\pi/q_{\textrm{min}}\approx 1600$~\AA. In Fig.~\ref{fig:magnetisation}, we show the temperature dependence of the intensity for selected $q$ below $q^{\ast}$. Near to $T_\textrm{C}$ it evolves as $M^2(T)\propto(1-\frac{T}{T_\textrm{C}})^{2\beta}$. We find that $\beta=0.32(1)$ describes our data perfectly in agreement with $\beta^{\textrm{theo}}=0.32$ for a three-dimensional (3D) Ising system~\cite{Chaikin}. This is also in good agreement with $\beta=0.36(1)$ from neutron diffraction~\cite{Kernavanois:2005}.

\begin{figure}[th]
	\begin{center}
		\includegraphics[width=0.85\columnwidth]{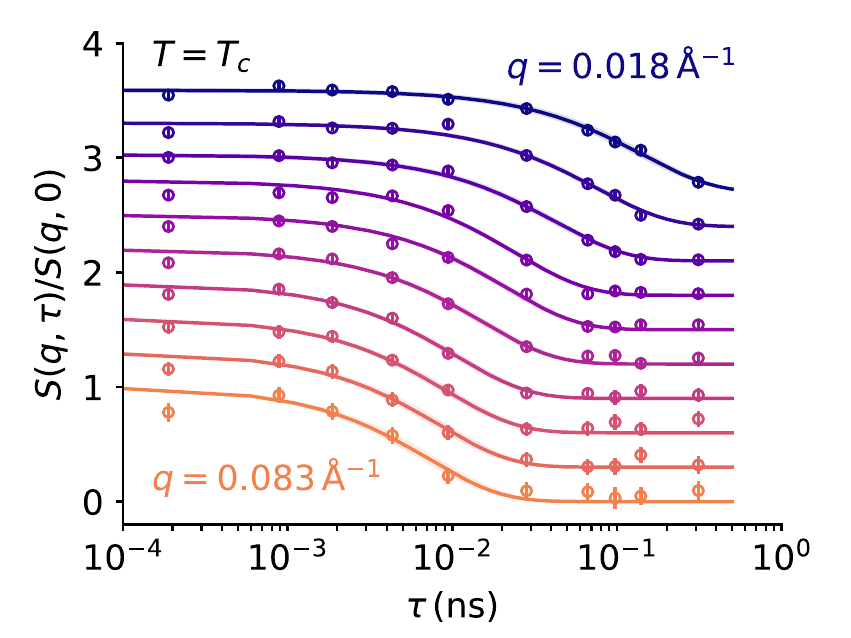}
		\caption{Spin fluctuation spectrum of \ug\ obtained by MIEZE. The intermediate scattering function $S(q,\tau)$ normalized to $S(q,0)$ (static signal) is shown at $T_\textrm{C}=52.7$~K for a range of momentum transfers $q$. Solid lines are fits to Eq.~\ref{eq:mieze_fit}.} 
		\label{fig:mieze}
	\end{center}
\end{figure}

For $q\geq q^{\ast}$ and for $T\approx T_\textrm{C}$ the $q$-dependence is described by a Lorentzian line shape characteristic of critical spin fluctuations with a correlation length $\xi$. The corresponding dynamical magnetic susceptibility is 
\begin{eqnarray}
	\frac{\chi''(\mathbf{q}, \omega)}{\omega}&=&\chi(\mathbf{q})\frac{\Gamma_q}{\Gamma_q^2+\omega^2}\label{Eq:dynamical_susceptibility}\\
	\chi(\mathbf{q})&=&\frac{\chi_0}{1+(\xi q)^2}\label{Eq:q-dependence},
\end{eqnarray}
where $\Gamma_q$ and $\chi_0$ are the momentum dependent relaxation frequency and the static magnetic susceptibility, respectively. Because of the longitudinal character of the spin fluctuations only $\chi_{aa}''$ is non-zero, and we have thus dropped the indices $i,j$. To investigate the critical scattering quantitatively, we subtract the Porod scattering [Fig.~\ref{fig:porod-and-lorentzian}] from the observed intensities [Fig.~\ref{fig:setup-and-energy-integrated}(c)]. For our experimental conditions the quasi-static approximation \cite{Marshall:1968, Squires:1978} is valid and thus integrating Eq.~\ref{Eq:cross-section} with respect to \hw, we obtain $\frac{d\sigma}{d\Omega}\propto T\chi(\mathbf{q})$(see supplemental material~\cite{supplement}). We show $\chi(\mathbf{q})$ obtained by dividing the observed intensity by $T$ for various temperatures in Fig.~\ref{fig:lorentzian}(b). The solid lines are fits to Eq.~\ref{Eq:q-dependence} to extract the $T$-dependence of $\chi_0$ and $\xi$ shown in Figs.~\ref{fig:susceptibility-correlation-length}(a) and (b), respectively. For comparison we show the static magnetic susceptibility $H/M$ determined by bulk magnetization measurements in Fig.~\ref{fig:susceptibility-correlation-length}(a) (blue squares) that scales perfectly with $\chi_0$.

We find that $1/\chi_0\propto (1-T/T_\textrm{C})^\gamma$ with $\gamma=1.23(3)$ and $1/\xi = \kappa = \kappa_0 (1-T/T_\textrm{C})^\nu$ with $\kappa_0=0.31(2)$\AA$^{-1}$ and  $\nu=0.63(2)$ [solid lines Figs.~\ref{fig:susceptibility-correlation-length}(a) and (b)]. The critical exponents are in excellent agreement with a 3D Ising FM, for which $\gamma^{\textrm{theo}}=1.24$ and $\nu^{\textrm{theo}}=0.63$~\cite{Chaikin}. Huxley {\it et al.} found $\kappa_0=0.29~$\AA$^{-1}$ in good agreement with our result. In contrast, they determined $\nu=1/2$, consistent with a mean-field transition~\cite{Huxley:2003}. However, their study was limited to $q>0.03$~\AA$^{-1}$ and underestimate the divergence of the critical fluctuations.

\begin{figure}[th]
    \begin{center}
        \includegraphics[width=0.9\columnwidth]{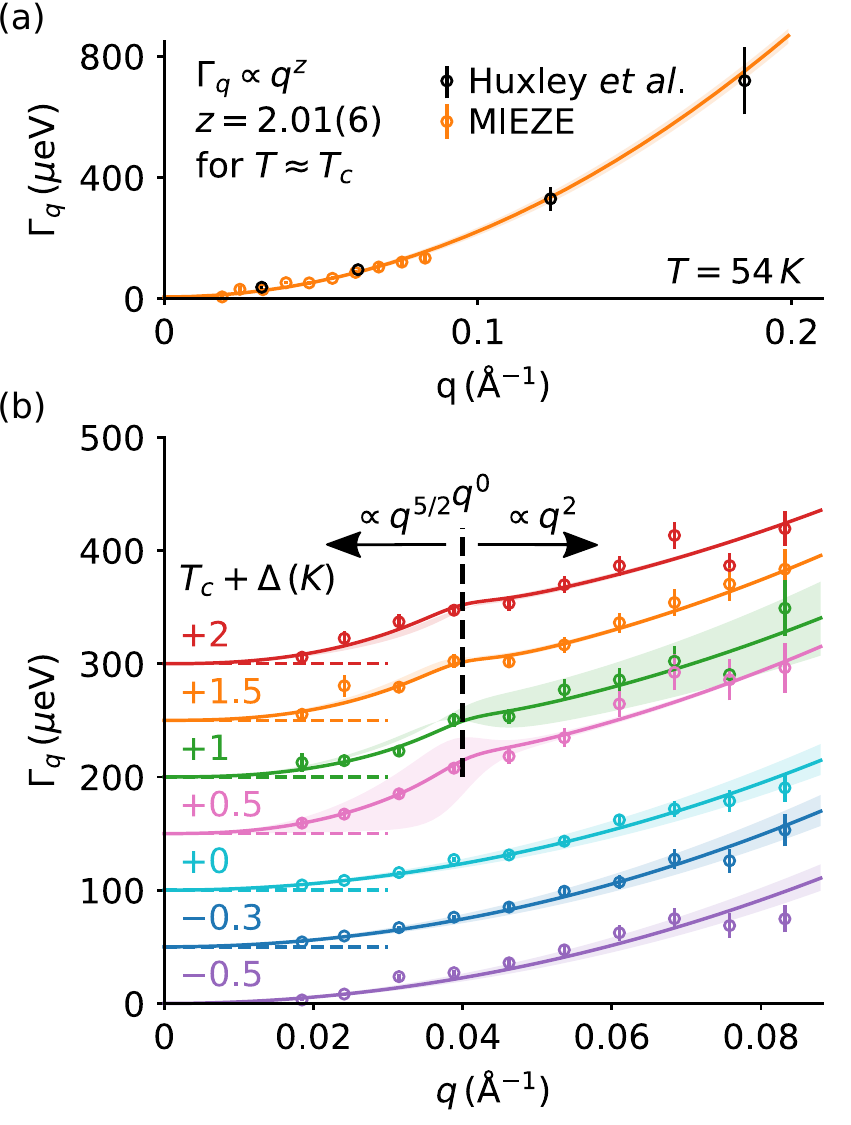}
		\caption{The fluctuation frequency $\Gamma_q$ of \ug\ at various temperatures $T$ as determined by fits of \ref{eq:mieze_fit} to the data shown in Fig.\ref{fig:mieze}. Solid lines are fits to $\Gamma_q\propto q^z$, where $z$ is the dynamical critical exponent. (a) Comparison of our data to the high-$q$ data by Huxley {\it et al.} \cite{Huxley:2003} is shown. (b) We find two distinct regimes with $z=2.5$ and 2 below and above $q^0=0.038$~\AA$^{-1}$, respectively (see text). Data sets are shifted by $50$~$\mu$eV for better readability as indicated by the horizontal dashed lines.}
        \label{fig:q-dependence-energy}
    \end{center}
\end{figure}

We now discuss the results of our MIEZE measurements. MIEZE measures the intermediate scattering function $S(q,\tau)$ that is the time Fourier transform of the scattering function $S(q,\omega)=1/\pi[n(\omega)+1]\chi_{ij}''(\mathbf{q}, \omega)$ (cf. Eq.~\ref{Eq:cross-section}). In Fig.~\ref{fig:mieze} we show $S(q,\tau)$ for various $q$ at $T_\textrm{C}$. $S(q,\tau)$ for all other measured temperatures are shown in Ref.~\onlinecite{supplement}. Because the spin fluctuations have Lorentzian lineshape (see Eq.~\ref{Eq:dynamical_susceptibility}) we fit $S(q,\tau)$ with an exponential decay [solid lines in Fig.~\ref{fig:mieze}]:
\begin{equation}
	S(q,\tau) = \exp(-|\Gamma_q| \cdot \tau).
	\label{eq:mieze_fit}
\end{equation} 	
The resulting fluctuation frequency $\Gamma_q$ is shown in Fig.~\ref{fig:q-dependence-energy}.

\begin{figure}[th]
    \begin{center}
        \includegraphics[width=0.9\columnwidth]{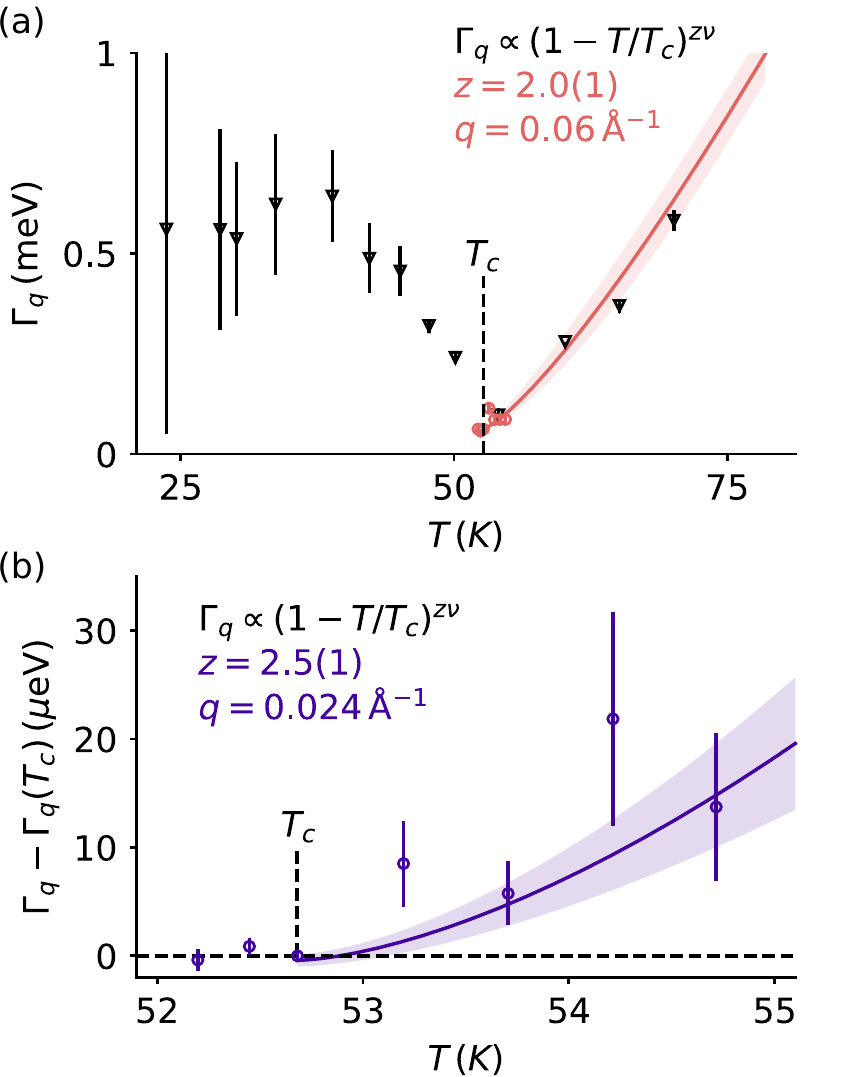}
        \caption{Temperature($T$)-dependence of the fluctuation frequency $\Gamma_q$ for $q$ above (a) and below (b) $q^0$. (a) Comparison to data of Ref~\onlinecite{Huxley:2003}. Solid lines denote $\Gamma_q\propto(1-T/T_\textrm{C})^{z\nu}$ (see text).}
        \label{fig:T-dependence-energy}
    \end{center}
\end{figure}

\begin{figure}[th]
    \begin{center}
        \includegraphics[width=0.9\columnwidth]{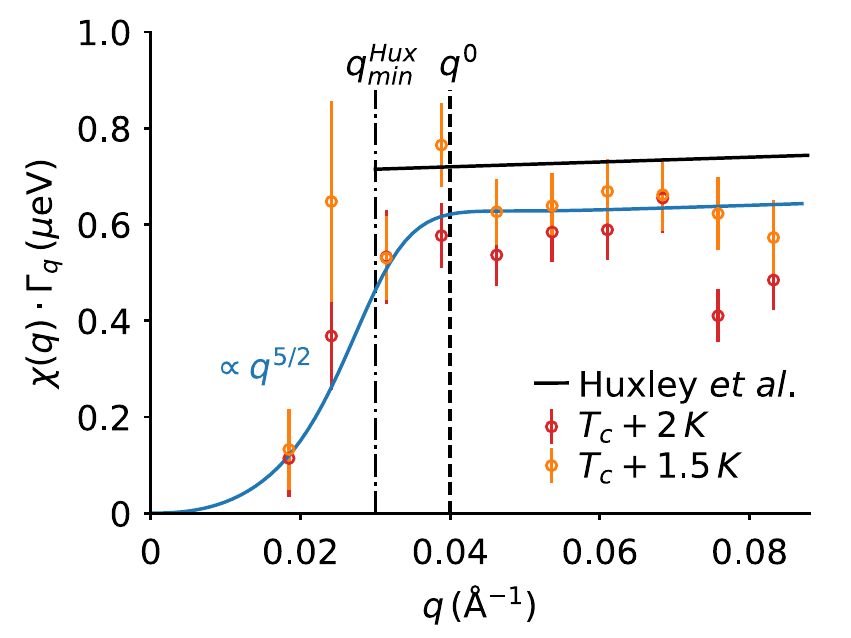}
        \caption{Temperature($T$)-dependence of the product of the magnetic susceptibility with the fluctuation frequency, $\chi(q)\Gamma_q$. The black line is $\chi(q)\Gamma_q$ as determined in Ref.~\onlinecite{Huxley:2003} that reports measurements down to $q^{\textrm{Hux}}_{\textrm{min}}=0.03$~\AA$^{-1}$ denoted by the dashed-dotted line. The blue line is a guide to the eye.}
        \label{fig:T-dependence-susceptibility-energy}
    \end{center}
\end{figure}

The momentum dependence of $\Gamma_q$ is described by the dynamical exponent $z$ via $\Gamma_q\propto q^z$. For $T\leq T_\textrm{C}$, we find that $\Gamma_q$ is fitted perfectly by $z=2.0(1)$ [Fig.~\ref{fig:q-dependence-energy}(a)]. This is in excellent agreement with predictions for a 3D Ising FM, for which $z^{\textrm{theo}}=2$~\cite{Chaikin}. For $T>T_\textrm{C}$, $\Gamma_q$ is also well described by $z=2$, however, only above a crossover value of $q^0$~=~0.04~\AA$^{-1}$. Below $q^0$, our data is best fit by $\Gamma_q=Aq^z$ with $z=2.53(4)$ (see Fig.~\ref{fig:q-dependence-energy}). This is consistent with $z$ = $5/2$ calculated for itinerant FMs within critical renormalization group theory \cite{Hohenberg:1977} and confirmed for various $d$-electron FMs such as Fe~\cite{Kindervater:2017}, Ni~\cite{Mirkiewicz:1969} and Co~\cite{Glitzka:1977}. Notably, typical values reported for $A$ are 3-350~meV\AA$^{5/2}$~\cite{Mirkiewicz:1969, Glitzka:1977, Dietrich:1976, Kindervater:2017} consistent with $A$=200(2)~meV\AA$^{5/2}$ that we find for \ug. As demonstrated in the Fig.~\ref{fig:q-dependence-energy}(a) for $T=54$~K, the fit of $\Gamma_q$ with $z=2.0(1)$ also describes the data of Huxley {\it et al.}~\cite{Huxley:2003} (black empty circles) perfectly. However, they conclude that $\Gamma_q$ remains finite for $q\longrightarrow 0$ in contrast to our findings. This discrepancy is easily explained by considering that their experiment was limited to $q\geq0.03$~\AA$^{-1}$, which is only slightly below $q^0$ where we observe the crossover to $z=5/2$.

Fig.~\ref{fig:T-dependence-energy} shows the $T$-dependence of $\Gamma_q$. For finite $q$, it follows the $T$-dependence of $\xi$ via $\Gamma_q\propto(1/\xi)^z=(1-T/T_\textrm{C})^{z\nu}$ in agreement with the dynamical scaling prediction~\cite{Halperin:1969}. In Fig.~\ref{fig:T-dependence-energy}(a) we show that for $q=0.06$~\AA$^{-1}$ both the results from Ref.~\onlinecite{Huxley:2003} and our own are consistent with $z=2$. Below $q^0$, $z=5/2$ agrees well with our data (solid line) consistent with the fits of $\Gamma_q$ shown in Fig.~\ref{fig:q-dependence-energy}.

For clean itinerant FMs the fluctuation spectrum is characterized by Landau damping as has been demonstrated for $3d$ transition metal materials~\cite{Lonzarich:1986, Bernhoeft:1988}. Here the product of the magnetic susceptibility with the fluctuation frequency, $\chi(q)\Gamma_q$, is given by the Lindhard dependence $(2/\pi)v_F\chi_P q$ for $T>T_\textrm{C}$, where $v_F$ and $\chi_P$ are the Fermi velocity and the non-interacting Pauli susceptibility, respectively \cite{Lonzarich:1985, Lonzarich:1999}. We show $\chi(q)\Gamma_q$ for \ug\ in Fig.~\ref{fig:T-dependence-susceptibility-energy}. Huxley {\it et al.}~\cite{Huxley:2003} who carried out measurements for $q\geq0.03$~\AA$^{-1}$ found that $\chi(q)\Gamma_q$ only weakly depends on $q$ and concluded that it remains finite for $q\longrightarrow 0$ [solid black line in Fig.~\ref{fig:T-dependence-susceptibility-energy}]. This difference with respect to prototypical $3d$ electron itinerant FMs is likely due to strong spin-orbit coupling that modifies the spin fluctuation spectrum. Our data agrees with the weak $q$ dependence above $q^0$ but clearly shows that $\chi(q)\Gamma_q\longrightarrow 0$ for $q\longrightarrow 0$, implying that the uniform magnetization is a conserved quantity in \ug. Our data is consistent with $\chi(q)\Gamma_q\propto q^{5/2}$ [solid blue line in Fig.~\ref{fig:T-dependence-susceptibility-energy}]. This more pronounced $q$-dependence is expected by theory near $T_\textrm{C}$~\cite{Lonzarich:1999}, and agrees with $\Gamma_q\propto q^{5/2}$. Here, we highlight that although the $q$-range over which $q^{z}$ with $z=5/2$ is observed is limited, this behavior is corroborated via three independent methods that are illustrated in Figs.~\ref{fig:q-dependence-energy}-\ref{fig:T-dependence-susceptibility-energy}.

\section{Summary}

Our results demonstrate that the spin fluctuations in \ug\ exhibit a dual character associated with  localized $5f$ electrons that are hybridized with itinerant $d$ electrons. Notably, as expected for a local moment FM with substantial uniaxial magnetic anisotropy all critical exponents determined from our results are in perfect agreement with the 3D Ising universality class~\cite{Chaikin}. Further, $\chi(q)\Gamma_q$ is approximately constant as a function of $q$ down to $q^0$ highlighting that the underlying spin fluctuations are localized in real space. In contrast, the dynamical exponent $z=5/2$ and $\chi(q)\Gamma_q\longrightarrow 0$ for $q\longrightarrow 0$ observed below the crossover value $q^0$, are characteristic of itinerant spin fluctations. Because the contribution of the conduction electrons to the total ordered moment is less than 3\% \cite{Kernavanois:2005}, below $T_\textrm{C}$ fluctuations of localized $f$ magnetic moments are dominant. Spin fluctuations with a dual character are consistent with the moderately enhanced Sommerfeld coefficient $\gamma=34$~mJ/K$^2$ mol of \ug~\cite{Lashley:2006, Troc:2012} and a next-nearest-neighbor uranium distance $d_{\textrm{U-U}}=3.85$~\AA~\cite{Oikawa:1996} near to the Hill value of 3.5~\AA~\cite{Hill:1970} that both suggest that the $5f$ electrons in \ug\ are hybridized with the conduction electrons.

In conclusion, the dual nature of spin fluctuations revealed by our MIEZE measurements strongly supports the scenario of p-wave superconductivity in \ug. First, to promote strong longitudinal fluctuations requires strong Ising anisotropy that typically is a result of localized $f$ electrons with substantial spin-orbit coupling, and is consistent with critical Ising exponents that we observe above $q^0$. Second, the theory for p-wave pairing assumes that it is the {\it same} itinerant electrons that are responsible for the coexisting FM and superconducting states~\cite{Fay:1980}, highlighting that the low-energy itinerant spin fluctuations below $q^0$ discovered here are crucial to mediate p-wave superconductivity. The maximum superconducting critical temperature $T_s$ occurs at the QPT at \px~\cite{Saxena:2001, Pfleiderer:2002}. Here a substantial increase of the Sommerfeld coefficient \cite{Tateiwa:2004} and changes in the electronic structure observed near $p_x$ \cite{Terashima:2001, Settai:2002} suggest that the hybridization of $5f$ electrons and conduction electrons increases at $p_x$ and corroborates that spin fluctuations with a dual nature are relevant for p-wave superconductivity. This is supported by a theory based on competition of FM exchange and the Kondo interaction that results in a localized to itinerant transition at $p_x$ \cite{Thomas:2011, Hoshino:2013}. 

Further, we note that our findings of longitudinal critical fluctuations in \ug\ are also consistent with the findings for UCoGe\cite{Hattori:2012}, which is another material that is a candidate for p-wave superconductivity. However, the results on UCoGe by Hattori {\it et al.}\cite{Hattori:2012} were obtained by NMR measurements that are unable to probe spin fluctuations at finite $q$ and, in turn, are unable to observe an intinerant-to-localized crossover as we report it here. Similarly, TAS measurements of UCoGe by Stock {\it et al.}\cite{Stock:2011} lack the required momentum and energy transfer resolution. 

Finally, we note that the crossover value $q^0$ corresponds to a length scale of approximately 160~\AA. The superconducting coherence length of \ug\ was estimated as $\xi^{\textrm{SC}}=200$~\AA~\cite{Saxena:2001}, which shows that the spin fluctuations relevant to the p-wave pairing are present at $q<q^0$. This may explain why triple-axis measurements of the spin fluctuation near $p_x$ with limited resolution were inconclusive~\cite{Kepa:2014}. Although, the pressure dependence of the crossover length scale $q^0$ remains to be determined to unambiguously associate it with the unconventional superconducting state in \ug, our results highlight that recent developments in ultra-high resolution neutron spectroscopy are critical for the study of low-energy spin fluctuations that are believed to drive the emergence of quantum matter states. Here the fluctuations that appear at zero $q$ such as for ferromagnetic and electronic-nematic quantum states can immediately be investigated via the MIEZE SANS configuration used here. In addition, MIEZE can be extended in straightforward fashion to study quantum fluctuations arising at large $q$~\cite{Martin:2018}, allowing for insights in antiferromagnetic QPTs and topological forms of order.   

\begin{acknowledgments}
The authors wish to thank the technical staff at MLZ for their help in conducting the experiments. We are grateful to Olaf Soltwedel and Bj{\"o}rn Pedersen for assistance with the MIEZE and neutron Laue diffraction measurements, respectively. We also acknowledge useful discussions with J. M. Lawrence, F. Ronning, and P. B\"{o}ni. Work at Los Alamos National Laboratory (LANL) was supported by LANL Laboratory Directed Research and Development program. Work at Technische Universit\"{a}t M\"{u}nchen was supported by the TRR80 Project F2. The authors acknowledge the financial support by the Federal Ministry of Education and Research of Germany in the framework of 'Longitudinale Resonante Neutronen Spin-Echo Spektroskopie mit Extremer Energie-Aufl\"osung' (project number 05K16WO6). The work by FH and MJ was supported through a Hans Fischer fellowship of the Technische Universit\"{a}t M\"{u}nchen --- Institute for Advanced Study, funded by the German Excellence Initiative and the European Union Seventh Framework Programme under grant agreement n$^{\circ}$ 291763. CP acknowledges financial support through ERC AdG ExQuiSid (788031) and DGF FOR960 (project 4). We also acknowledge support from the European Union through the Marie-Curie COFUND program.
\end{acknowledgments}

\end{bibunit}
\newpage
\begin{bibunit}
\balancecolsandclearpage
\renewcommand{\thefigure}{S\arabic{figure}}
\renewcommand{\theequation}{S\arabic{equation}}
\renewcommand{\thetable}{S\arabic{table}}

\section*{SUPPLEMENTAL MATERIAL}

\setcounter{section}{0}
\setcounter{figure}{0}
\setcounter{table}{0}

In this supplemental material we describe the characterization of the \ug\ sample with neutron Laue diffraction and neutron depolarization analysis, as well as detailed information about the used MIEZE setup, and additional SANS measurements. We also describe certain approximation for quasi-elastic scattering at high temperature as well as resolution calculations that demonstrate that all observed spin fluctuations are purely longitudinal in nature.

\vskip2pc

\maketitle

\section{Neutron Laue Diffraction}
\begin{figure}[th]
\begin{center}
    \includegraphics[width=0.8\columnwidth]{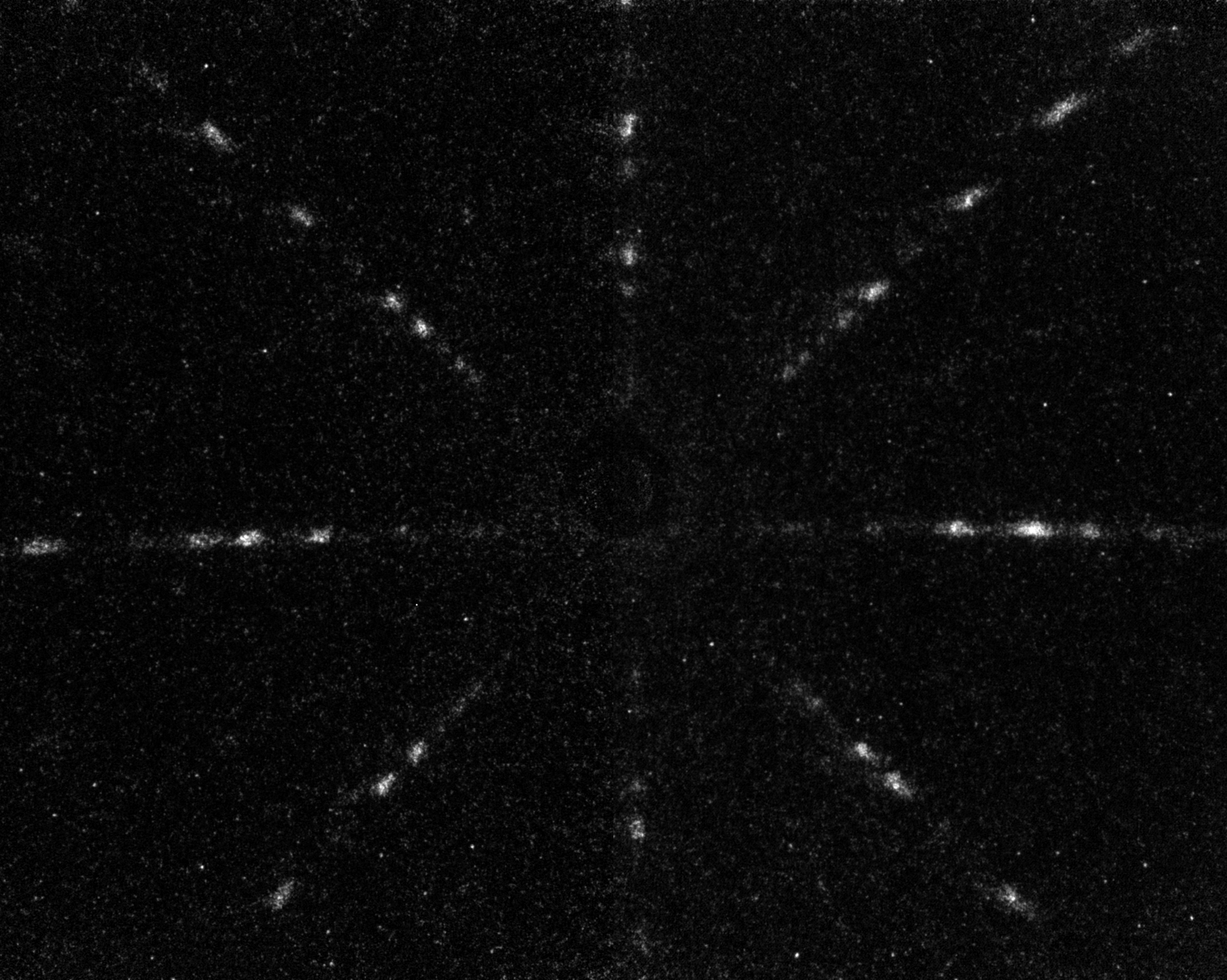}
    \caption{Neutron Laue image recorded at the nLaue instrument at MLZ with the crystallographic $b$ axis approximately parallel to the incident neutron beam.}
    \label{fig:Laue}
\end{center}
\end{figure}

In order to demonstrate the high quality of the \ug\ single crystal used for the experiments in the main text, as well as to orient it precisely, we have carried out a neutron Laue diffraction experiment using the nLaue instrument at the MLZ. The nLaue instrument uses a white thermal beam, where a sapphire filter is installed in the shielding to avoid background from fast neutrons and gamma radiation. This results in a neutron spectrum with suitable intensity between 0.8 and 4 \AA. The generated Laue diffraction patterns are recorded with a Photonic Sciene position-sensitive detector with a large active area of 252 x 198 mm$^2$ and a resolution of 2088 x 1554 pixels based on a $^{6}$Li doped ZnS scintillator with two cooled CCD-cameras with channel plate intensifiers. Our experiment was carried out in Laue backscattering geometry. A representative Laue pattern from the sample used for the SANS and MIEZE measurements is shown in Fig.~\ref{fig:Laue} and demonstrates that there is only a single grain.

\section{Neutron Depolarization Imaging}

We employed neutron depolarization imaging (NDI) measurements \cite{Schulz:2010} to establish that the \ug\ single crystal used for our SANS and MIEZE measurements is magnetically homogeneous. We note that for experiments attempting to extract critical exponents this is an often ignored requirement. Notably, it has been demonstrated previously that for large ferromagnetic samples required for neutron spectroscopy, the magnetic properties are often inhomogeneous, where particularly the Curie temperature $T_\textrm{C}$ has been shown to vary more than 20 K over the entire sample \cite{Schulz:2010}.

The NDI technique is based on the combination of a neutron imaging beam line using a position sensitive detector with a neutron polarization analysis setup. It allows to resolve the influence of a sample on the neutron polarization spatially. Notably, for a ferromagnet cooled below the Curie temperature $T_\textrm{C}$, any component of the neutron polarization that is perpendicular to the magnetization direction will start to precess, in turn, resulting in partial depolarization of the beam. In turn, this allows for a spatially resolved measurement of $T_\textrm{C}$. 

The NDI measurements carried out on the \ug\ single crystal studied here were carried out at the imaging beam line ANTARES at MLZ. Except for the detector the setup used here is identical to the one described in Ref.~\cite{Seifert:2017}. For this experiment the detector setup consisted of a 200 $\mu$m LiF/ZnS scintillator to convert the neutron beam into visible light, magnifying optics with a magnification of 1:5.6, and finally an Andor iKon-L camera with 13.5 $\mu$ pixels. The effective pixel size shown in this setup amounts to 75 $\mu$m. The effective spatial resolution of this setup is about 1~mm as determined by the distance between the sample and the detector of 500~mm and the $L/D$ ratio, where $D$ is the size of the pinhole aperture and $L$ is the distance between the aperture and the detector. Here we used $L/D=500$.

\begin{figure}[th]
\begin{center}
    \includegraphics[width=0.9\columnwidth]{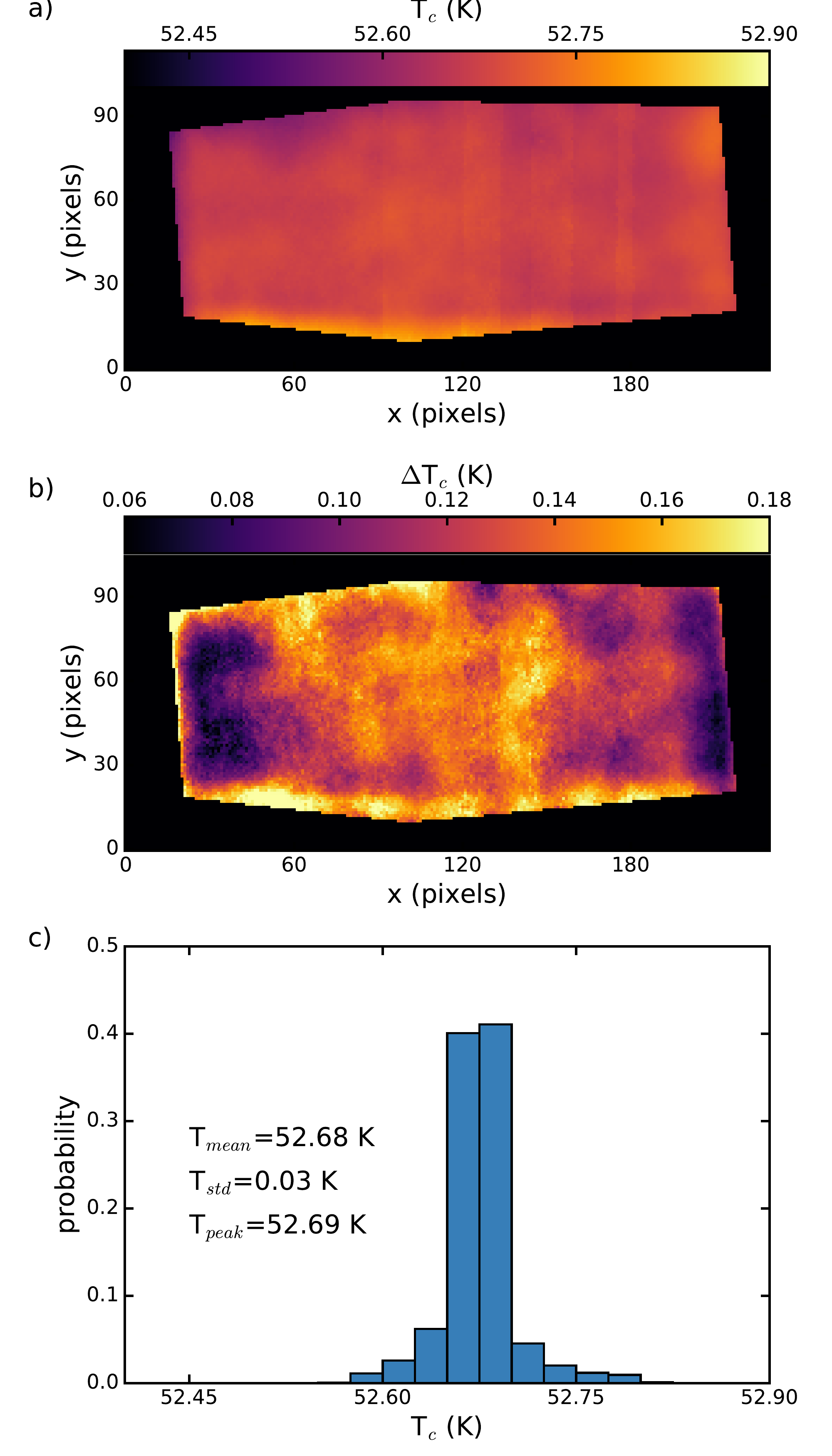}
    \caption{Results of the neutron depolarization imaging (NDI) measurements of the \ug\ single crystal used for our study. (a) The spatial distribution of the Curie temperature $T_\textrm{C}$ is shown over the entire single crystal. (b) The spatial variation $\Delta T_\textrm{C}$ is shown. (c) Histogram of the probability of various $T_\textrm{C}$ across the sample is shown.}
    \label{fig:NDI}
\end{center}
\end{figure}

In Fig.~\ref{fig:NDI}, we summarize the analyzed results of our NDI measurements on \ug. In Fig.~\ref{fig:NDI}(a) the spatial distribution of the Curie temperature $T_\textrm{C}$ is shown across the sample. Here the $c$ axis is approximately along the long axis of the sample. In Fig.~\ref{fig:NDI}(b) shows the variation $\Delta T_\textrm{C}$ across the sample. We have used the information from Fig.~\ref{fig:NDI}(a) to compile a histogram for the probability of various $T_\textrm{C}$ observed across the sample. The average  $T_\textrm{C}$ = 52.68 K and we find a standard deviation of 0.03 K, demonstrating that the sample used for the measurements reported here is magnetically entirely homogeneous, and thus ideal for the investigation of critical magnetic properties. 

\section{SANS measurements}

We performed supporting small angle neutron scattering (SANS) experiments at the instrument SANS-1\cite{Muehlbauer:2016} at MLZ in order to verify the Porod scattering due to ferromagnetic domains in \ug\ to smaller momentum transfers. The experiment was carried out with a neutron wavelength of $\lambda$ = 6.5~\AA~ selected via a velocity selector. The sample was oriented identical to the MIEZE measurements. A sample-detector distance of 20~m in combination with the detector being asymmetrically moved on one side of the direct beam position allowed for a momentum transfer range $q$ = 0.002-0.02~\AA$^{-1}$. Together with the $^3$He area detector with 8\,x\,8\,mm resolution, a momentum transfer resolution of $\Delta q \approx 8\cdot 10^{-4}\,$\AA$^{-1}$ was achieved. The data was collected by performing rocking scans at each temperature. Background data recorded well above $T_\textrm{C}$ was subtracted from all data sets shown, to remove non-magnetic scattering. The momentum transfer and temperature dependence of the magnetic scattering below $T_\textrm{C}$ obtained in this way is shown in Fig.~2 in the main text (square symbols).

\section{Approximations for energy-integrated critical scattering}

Here we show how the measured neutron cross-section for inelastic magnetic scattering provided in Eq.~1 in the main text can by simplified for $\hbar\omega\ll k_B T$. Via a Taylor expansion in $x=\frac{\hbar\omega}{k_B T}$ it follows that $\frac{1}{1-e^{-\frac{\hbar\omega}{k_B T}}}\approx\frac{k_B T}{\hbar\omega}$. For this case the scattering function simplifies to
\begin{equation}
	S(\mathbf{q},\omega)=\frac{\chi_{\alpha\beta}''(\mathbf{q}, \omega)}{1-e^{-\beta \hbar\omega}}\approx\chi_{\alpha\beta}''(\mathbf{q}, \omega)\frac{k_B T}{\hbar\omega}.\label{Eq:simple_S}
\end{equation}
All data collection for our neutron scattering study in \ug\ was carried out near $T_\textrm{C}=52.7 K$, which corresponds to $k_B T_\textrm{C}=4.5$~meV. In addition, it is already known from previous triple axis spectroscopy measurements that $\hbar\Gamma_q<\hbar\Gamma^{\textrm{max}}=0.3$~meV for the $q$-range up to 0.08~\AA$^{-1}$ that we have investigated here \cite{Huxley:2003}. Our own measurements also confirm this (cf. Fig.~7 in the main text). This means that for the relevant energy range $\frac{\hbar\Gamma^{\textrm{max}}}{k_B T_\textrm{C}}\approx0.07\ll 1$.

Further, for the energy-integrated intensity recorded as function of temperature with the MIEZE setup disabled, we can make use of the so-called quasi-static approximation \cite{Marshall:1968, Squires:1978}, which says that when the energy of incident neutrons $E_i$ is larger than the relaxation frequency $\hbar\Gamma$ of the spin fluctuations that are being investigated the energy-integrated neutron scattering cross-section is given by
\begin{equation}
	\frac{d\sigma}{d\Omega}\propto (\delta_{\alpha\beta}-\hat{Q}_{\alpha}\hat{Q}_{\beta})|F_{\mathbf{Q}}|^2 S(\mathbf{q}, 0),\label{Eq:quasi-static-cross-section}
\end{equation}
where 
\begin{equation}
	S(\mathbf{q}, t)=\hbar\int S(\mathbf{q}, \omega)\exp(i\omega t) d\omega,\label{Eq:intermediate_scattering}
\end{equation}
is the intermediate scattering function, which is the Fourier transform of the scattering function $S(q,\omega)=1/\pi[n(\omega)+1]\chi_{ij}''(\mathbf{q}, \omega)$ [cf. Eq.~1 in the main text] with respect to time. The experiments on RESEDA were carried out with an incident wavelength $\lambda=6$~\AA~, which corresponds to  $E_i=2.3$~meV, and thus we have $\frac{\hbar\Gamma^{\textrm{max}}}{E_i}\approx0.1\ll 1$, and the quasi-static approximation is valid for our experiment. 

Using Eq.~2 from the main text and Eqs.~\ref{Eq:simple_S} and \ref{Eq:intermediate_scattering}, we obtain
\begin{equation}
	S(\mathbf{q}, t=0)=\int k_B T \chi(\mathbf{q})\frac{\Gamma_q}{\Gamma_q^2+\omega^2} d\omega=\pi k_B T \chi(\mathbf{q})\label{Eq:intermediate_final}.
\end{equation}
Taking into account that for the used SANS geometry $|F_{\mathbf{Q}}|^2\approx 1$, we find the energy-integrated neutron cross-section for the spin fluctuations in \ug\ as
\begin{eqnarray}
	\frac{d\sigma}{d\Omega}&\propto&\pi k_B T\chi(\mathbf{q})\nonumber\\
	&=&\pi\frac{k_B T\chi_0}{1+(\xi q)^2}.\label{Eq:energy-integrated-critical-scattering}
\end{eqnarray}
In conclusion, the static susceptibility $\chi_0$ and correlation length $\xi$ can be directly obtained by fitting the observed energy-integrated intensities with Eq.~\ref{Eq:energy-integrated-critical-scattering}.

\begin{table*}[th!]
    \caption{Instrument Parameters used to calculate the $q$ resolution of the instrument Reseda in energy-integrating mode.}\label{tab:resolution_parameters}
    \begin{tabular}{l   l   r   r}
        parameter							& unit & variable 		& value\\
		\hline
		detector pixel size					&(mm)			& $\Delta$		& 1.56 \\
        incoming wavelength                 &(\AA$^{-1}$) 	& $\lambda$ 	& $5.918$\\
        wavelength spread (FWHM)			& (1)           & $\Delta \lambda/\lambda$ & 0.117 \\
        scattering angles                   & (degrees)		& $2 \theta$	& 0.98 - 4.54\\
        source aperture horizontal (half width $\hat{=} $``radius'') & (mm) 	& $r_{1,h}$		& 5\\
        source aperture vertical            & (mm)	    	& $r_{1,v}$		& 10\\
        defining aperture horizontal        & (mm)      	& $r_{2,h}$		& 1.5\\
        defining aperture vertical          & (mm)      	& $r_{2,v}$		& 16\\
        source aperture  - defining aperture & (mm) & $L$   & 1450 \\
        distance defining aperture - detector &(mm)& $l$	& $\approx 2500$\\
        distance sample - detector 			& (mm)& $L_{SD}$		& 2230\\
        \hline
    \end{tabular}	
\end{table*}

\section{Resolution effects}

Here we show unambiguously that the critical spin fluctuations in \ug\ are purely of longitudinal character $\delta S_\parallel$ and that the observation of intensity in the configuration with the magnetic easy axis $a$ perpendicular to the direction of the incident neutron beam $\mathbf{n}$ that solely probes transverse spin fluctuations $\delta S_\perp$ is due to resolution effects. We note that here we purely consider the energy-integrated mode for which the MIEZE spectrometer was switched off and thus corresponds to a SANS experiment. For Reseda the \vQ\ resolution is mostly determined by two apertures installed upstream of the sample and can be calculated following Ref.~\onlinecite{Pederson:1990}. Here the first apertures is the source aperture at the front of the instrument and the second is a defining aperture installed directly in front of the sample. All relevant instrument parameters used for the following calculation are summarized in the Table~\ref{tab:resolution_parameters}. The \vQ-resolution has three different contributions that we consider in the following.

\begin{figure}[th]
	\begin{center}
		\includegraphics[width=0.9\columnwidth]{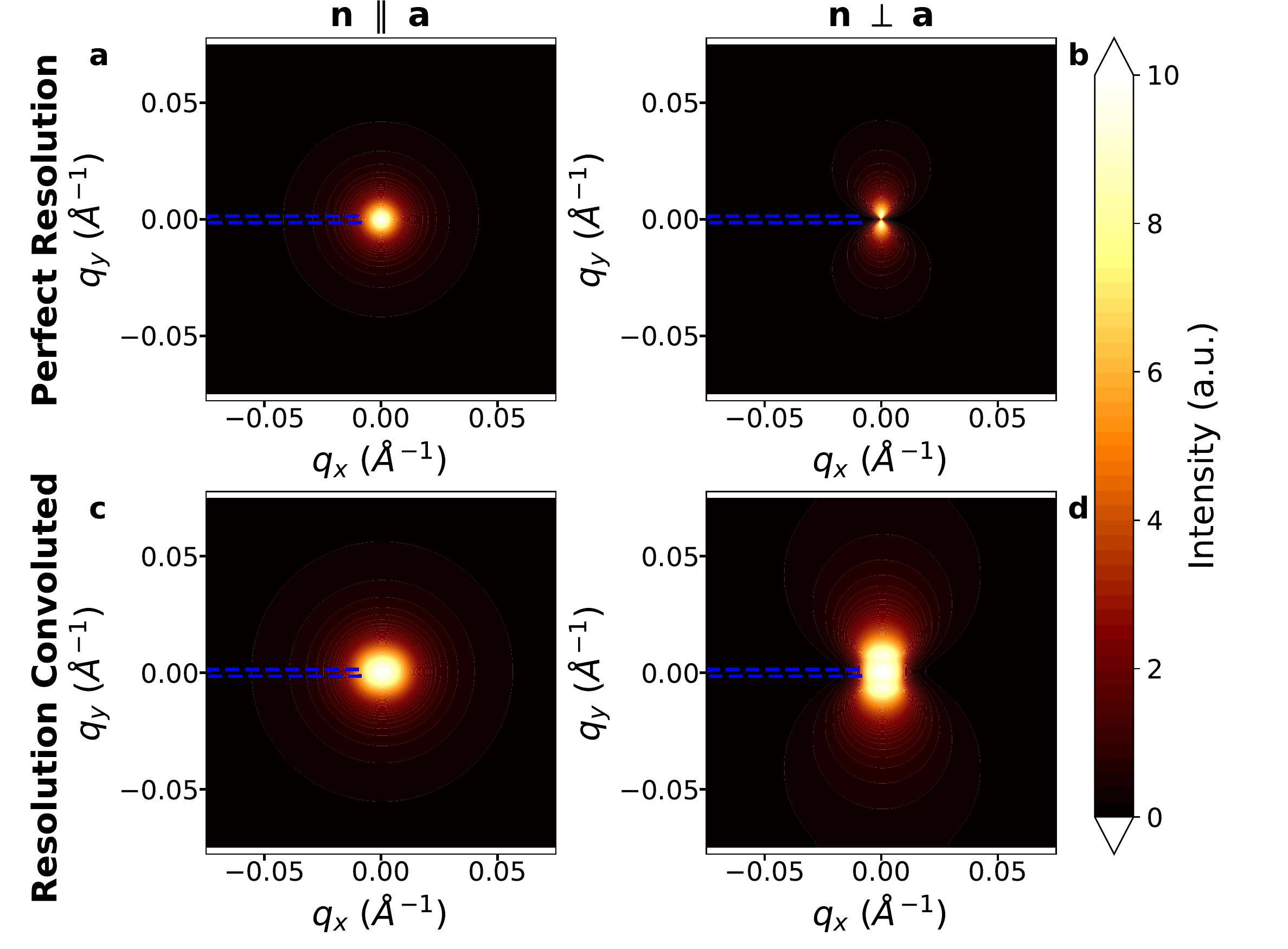}
		\caption{Simulation of the effects of instrument resolution on the observation of intensity associated with longitudinal critical spin fluctuations in \ug\ at $T$ = 52.8~K (see text for details. (a) and (b) show simulated intensity for the case of perfect resolution for the two configurations with the the crystallographic $a$-axis, which is the magnetic easy-axis for \ug\, oriented parallel ($\mathbf{n}\parallel a$) and perpendicular ($\mathbf{n}\perp a$) to the incident neutron beam, respectively. (c) and (d) show the same configurations as (a) and (b), respectively, however, the simulated intensities were convoluted with the instrument resolution. The blue region-of-interest denotes the $q$-cut that is plotted in Fig.~\ref{fig:resolution_temp}.}
		\label{fig:resolution_detector}
	\end{center}
	\end{figure}
	
	\begin{figure}[bh!]
	\begin{center}
		\includegraphics[width=0.9\columnwidth]{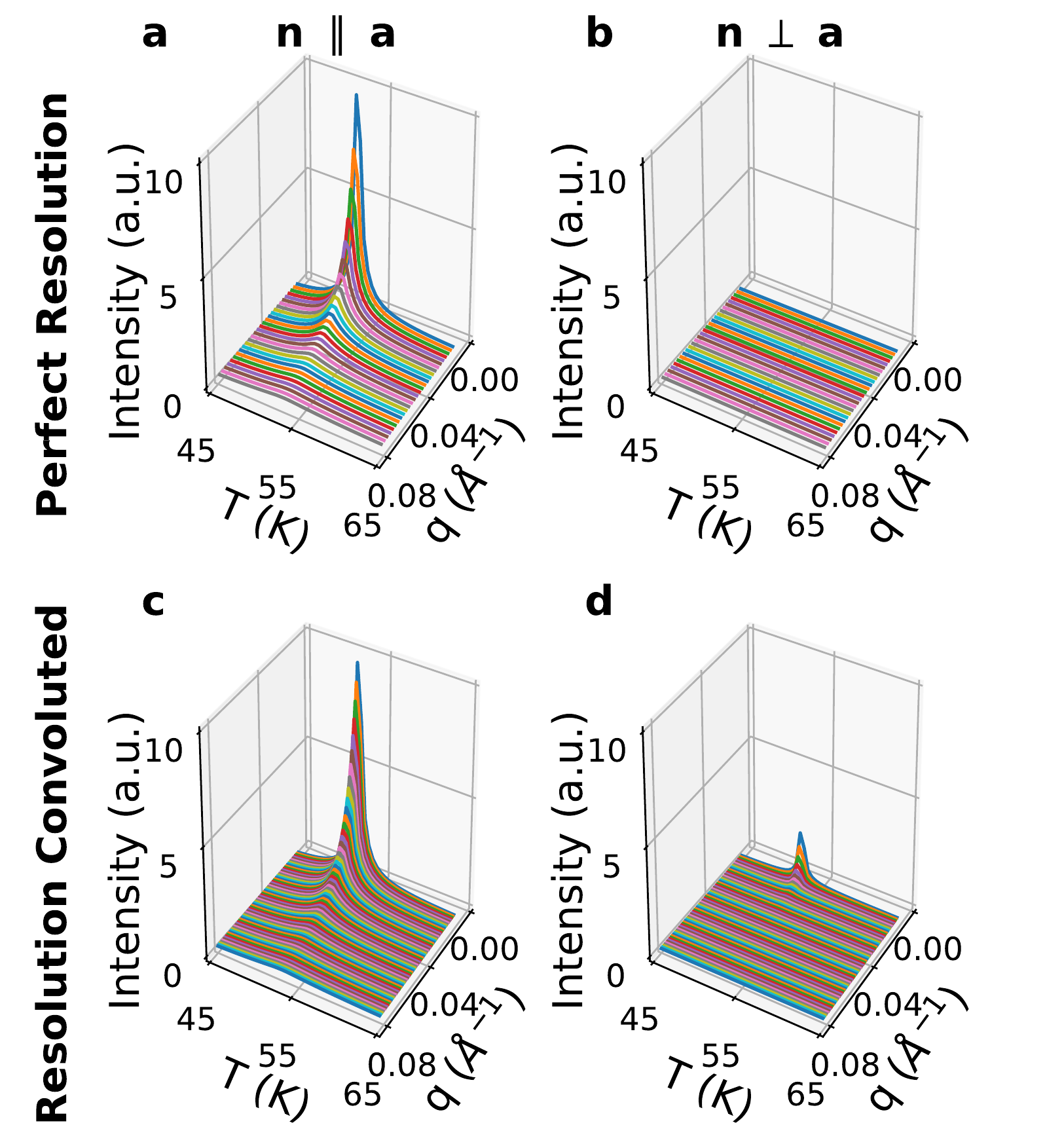}
		\caption{Simulation of the effects of instrument resolution on the observation of intensity associated with longitudinal critical spin fluctuations in \ug\ as a function of temperature $T$ (see text for details). The panels (a)-(d) show the $q$-cut denoted with the blue region-of-interest in Fig.~\ref{fig:resolution_detector}(a)-(d). Here (a) and (b) show results assuming perfect instrument resolution for the two configurations $\mathbf{n}\parallel a$ and $\mathbf{n}\perp a$, respectively (see text). (c) and (d) show results for the same configurations, however, the simulated intensity is folded with the instrument resolution corresponding to our experiments on \ug.
		}
		\label{fig:resolution_temp}
	\end{center}
\end{figure}

The first contribution is due to the wavelength spread $\frac{\Delta \lambda}{\langle \lambda \rangle}$ allowed by the selector, where $\langle \lambda \rangle$ is the selected incident wave length.
\begin{equation}
	\sigma_W = \sigma_{\lambda} \frac{\langle q\rangle}{\langle \lambda \rangle} = \langle q \rangle \frac{\Delta \lambda}{\langle \lambda \rangle} \frac{1}{2(2 \ln 2)^{1/2}}.
\end{equation}
The second contribution is given by the collimation spread due to the defining apertures in front of the sample. Within the scattering plane it is defined via
\begin{align}
	\sigma_{Col||q} = \frac{\langle q \rangle \cos \langle \theta \rangle \Delta \beta_1}{2(2 \ln 2)^{1/2}},
\end{align}
where $\langle q \rangle$ is the central momentum transfer and $\langle 2\theta \rangle$ is the scattering angle. $\Delta \beta_1$ is defined as
\begin{equation}
	\Delta \beta_1 = 2r_{1,h}/L - \frac{1}{2} \frac{r_{2,h}^2}{r_{1,h}}\frac{\cos^4\langle 2\theta \rangle}{l^2 L} \cdot (L + l/\cos^2\langle 2\theta \rangle)^2.
\end{equation}
Here $2r_{1,h}$ and $2r_{2,h}$ are the width of the source and the defining aperture, respectively. $L$ and $l$ are the distance between the source and defining apertures and the distance between the defining aperture and the detector respectively.

Similarly, the collimation spread perpendicular to the scattering plane is given by
\begin{equation}
	\sigma_{Col \perp q} = \frac{\langle q \rangle \Delta \beta_2}{(2 \ln 2)^{1/2}},
\end{equation}
where 
\begin{equation}
	\Delta \beta_2 = 2 r_{1,v}/L  - \frac{1}{2} \frac{r_{2,v}^2}{r_{1,v}} \frac{\cos^2 \langle 2 \theta \rangle}{l^2 L} \cdot (L + l/\cos \langle 2 \theta \rangle)^2.
\end{equation}
$2r_{1,v}$ and $2r_{2,v}$ are the height of the source and the defining aperture, respectively.

The last contribution is the detector resolution, where the resolution parallel and perpendicular to the scattering vector $q$ are given by
\begin{equation}
	\sigma_{Det||q} = \langle q \rangle \cos \langle \theta \rangle \cos^2 \langle 2\theta \rangle \Delta [l 2(2 \ln2)^{1/2}]^{-1},
\end{equation}
and
\begin{equation}
	\sigma_{Det \perp q} = \langle q \rangle \cos \langle 2\theta \rangle  \Delta [l (2 \ln2)^{1/2}]^{-1},
\end{equation}
respectively. $\Delta$ denotes the pixel size of the detector. 

The combined resolution by all three contributions is then
\begin{equation}
	\sigma^2 = \sigma_W^2 + \sigma_{Col}^2 + \sigma_{Det}^2. 
\end{equation}
Using the values for all parameters defined in Table~\ref{tab:resolution_parameters} we obtain the following instrument resolution. In the vertical direction the resolution $\Delta q_y=0.003$~\AA$^{-1}$ remains constant over the entire momentum transfer range $0.015\leq q \leq 0.083$~\AA$^{-1}$ observed in our  experiments. The resolution in the scattering plane varies smootly from $\Delta q_x=0.003$~\AA$^{-1}$ to $0.005$~\AA$^{-1}$ from the smallest to the largest momentum transfer. 

Using this information we can simulate how the resolution impacts our experiment. Using Eq.~\ref{Eq:energy-integrated-critical-scattering} and the parameters $\gamma=1.23$, $\nu=0.63$, $\kappa_0=0.31$~\AA$^{-1}$, and $T_\textrm{C}$ = 52.7 K we simulate the intensity of the energy-integrated critical spin fluctuations at the temperature $T=52.8$~K as observed on the position sensitive detector of RESEDA. In Fig.~\ref{fig:resolution_detector}(a) and (b), we show the results of this calculation for the two configurations with the the crystallographic $a$-axis, which is the magnetic easy-axis for \ug\, oriented parallel ($\mathbf{n}\parallel a$) and perpendicular ($\mathbf{n}\perp a$) to the incident neutron beam, respectively. As explained in the main text, for $\mathbf{n}\parallel a$ both transverse and longitudinal spin fluctuations can be observed, whereas for ($\mathbf{n}\perp a$) only transverse fluctuations can be observed. For this calculation we have assumed that the instrumental resolution is perfect and that the critical spin fluctuations are of purely longitudinal character. In Fig.~\ref{fig:resolution_detector}(c) and (d), we show the result of the same calculation, however, we have convoluted the signal of the spin fluctuations with the instrumental resolution calculated above. It is obvious that the resolution affects the result. We note that in the main text, we have not used entire detector images, but only plotted intensity along a trajectory in reciprocal space denoted by the blue region of interest in Fig.~\ref{fig:resolution_detector}. 

In Fig.~\ref{fig:resolution_temp}, we have repeated the calculation above, however for various temperatures $45\leq T\leq 65$~K. Here we only show the $q$-cut along the blue region of interest in Fig.~\ref{fig:resolution_detector} for each configurations. Fig.~\ref{fig:resolution_temp}(a) and (b) show the result for two configurations $\mathbf{n}\parallel a$ and $\mathbf{n}\perp a$, respectively, for the case of perfect instrument resolution. Because in the calculation we have assumend that the critical fluctuations are purely longitudinal, we don't see any intensity for $\mathbf{n}\perp a$ as shown in Fig.~\ref{fig:resolution_temp}(b). Fig.~\ref{fig:resolution_temp}(c) and (d) also show the result for $\mathbf{n}\parallel a$ and $\mathbf{n}\perp a$, respectively, however, this time the signal associated with the longitudinal spin fluctuations is convoluted with the instrument resolution of our experiment. It can be clearly seen that now there is a small amount of intensity visible for for $\mathbf{n}\perp a$ shown in Fig.~\ref{fig:resolution_temp}(d), illustrating that the instrumental resolution indeed introduces artefacts in the channel that is purely sensitive to transverse spin fluctuations. In conclusion, this shows unambiguously that the spin fluctations in \ug\ are purely longitudinal.
	
\begin{figure}[th]
	\begin{center}
		\includegraphics[width=3.3in]{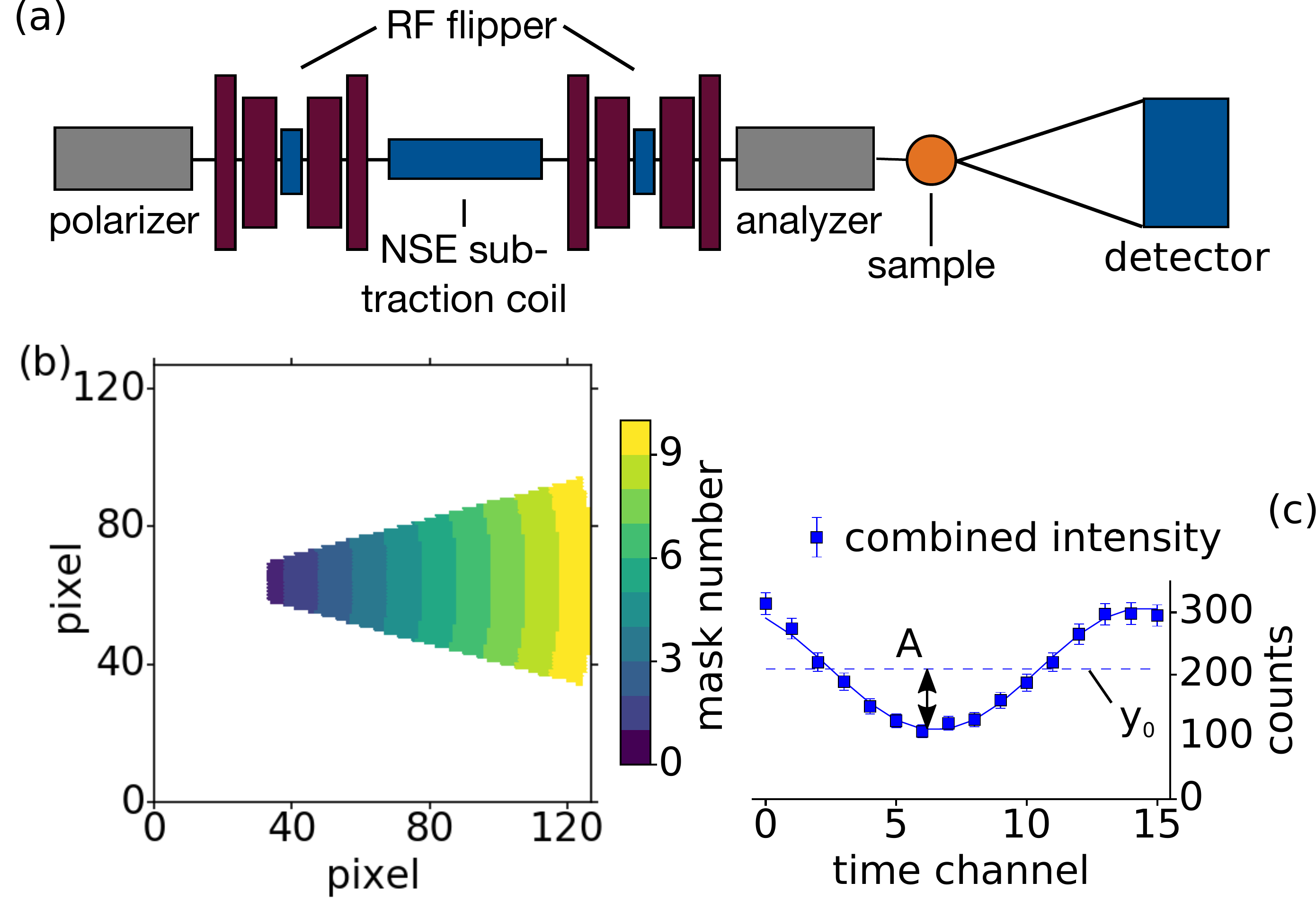}
		\caption{(a) MIEZE setup at the instrument RESEDA. The neutron flight path is from left to right. (b) Evaluated detector masks ranging from $q=0.018\,\text{\AA}^{-1}$ up to $q=0.083\,\text{\AA}^{-1}$. (c) Combined intensity after phase shifting the signals by predefined phase from the graphite measurement. Lines are fits to Eq.\,\ref{eq:sinus}.} 
		\label{Fig:MIEZE_setup}
	\end{center}
\end{figure}

\section{Details of MIEZE setup and data analysis}

For a detailed description of the MIEZE technique we refer to Keller et al\,\cite{2002_Keller_S}. Here we have employed a longitudinal MIEZE setup (all magnetic field axis are parallel to the neutron beam) as realised at RESEDA at the MLZ Garching as shown in Fig.\,\ref{Fig:MIEZE_setup}(a) \cite{2015_Franz_JLRFJ}. The neutron beam is polarized by a 2\,m long V-cavity polarizer. Two resonant spin flippers separated by a distance $L_1$ run at frequencies $\omega_{A,B} > 35\,$kHz where the second runs at a higher frequency $\omega_B > \omega_A$. In combination with the transmission bender as analyzer, a time dependent, sinusoidal intensity modulation with frequency $2\cdot(\omega_B - \omega_A)$ is produced at a focus point, $L_{SD}$ behind the sample, where a 2D position sensitive and time resolving CASCADE detector with high time resolution ($\Delta t$ = 50 nsec) and an active area of 200 x 200 mm$^2$\cite{Klein:2009, HAUSSLER:2011} is positioned. The amplitude of the modulated beam, the so-called MIEZE contrast, takes the role of the neutron polarization in neutron spin echo spectroscopy experiments. A field integral subtraction coil is used to reach spin echo times below $\sim 100$\,ps \cite{2005_Haussler_PCCP}. In our experiments a dynamic range from 0.2\,ps to 310\,ps was covered. 

The neutron beam of RESEDA is defined by a 20\,x\,10\,mm (height x width) aperature in front of the second RF flipper and a 32\,x3\,mm slit after the analyzer separated by 1450\,mm. A rectangular cadmium aperture directly in front of the sample minimized the background. The direct beam was blocked by a rectangular beam stop directly in front of the CASCADE detector. The pixels of the CASCADE detector have a size of 1.56\,x\,1.56\,mm. 

Fig.\,\ref{Fig:MIEZE_setup}(b) illustrates the detector masks that were used to bin the data in $q$. The masks are concentrical around the direct beam with an opening angle of $30^{\circ}$. The period of the sinusoidal intensity modulation is binned into 16 time channels. Typical data is shown in Fig.\,\ref{Fig:MIEZE_setup}(c). The phase depends on the flight path between the sample and the point of detection. It is predefined for every mask by means of a calibration measurement using an ideal elastic scatterer, in this case a graphite powder at room temperature. The data from individual foils is shifted by the phase correction and subsequently combined. The data is fitted by the function

\begin{equation}
 	y = A \cdot \sin(f \cdot t + \phi_0) + y_0
 	\label{eq:sinus}
 \end{equation} 
with the amplitude $A$, the mean-value $y_0$, the phase $\phi_0$ and the frequency $f$ which is fixed. The contrast is defined as $C = A/y_0$. After background subtraction the data is normalised to the signal of an elastic scatterer to account for the instrumental resolution. Here we used graphite. The contrast $C$ is directly proportional to the Fourier cosine transform of the scattering function $S(q,\omega)=1/\pi[n(\omega)+1]\chi_{ij}''(\mathbf{q}, \omega)$ [cf. Eq.~1 in the main text] with respect to time $\tau$. Note that because in this case $S(q,\omega)$ has symmetric Lorentzian line shape and is centered at $\hbar\omega=0$ [cf. Eq.~1 in the main text] the Fourier cosine transform is equivalent to a full Fourier transform and we obtain [cf. Eq.~\ref{Eq:intermediate_scattering}]:
\begin{equation}{}
	S(q,\tau) = \hbar \int S(q,\omega)  \cos(\omega \cdot \tau_{MIEZE}) d\omega. 
	\label{EQ:Fourier transform}
\end{equation}

\begin{figure}[th]
\begin{center}
	\includegraphics[width=3.3in]{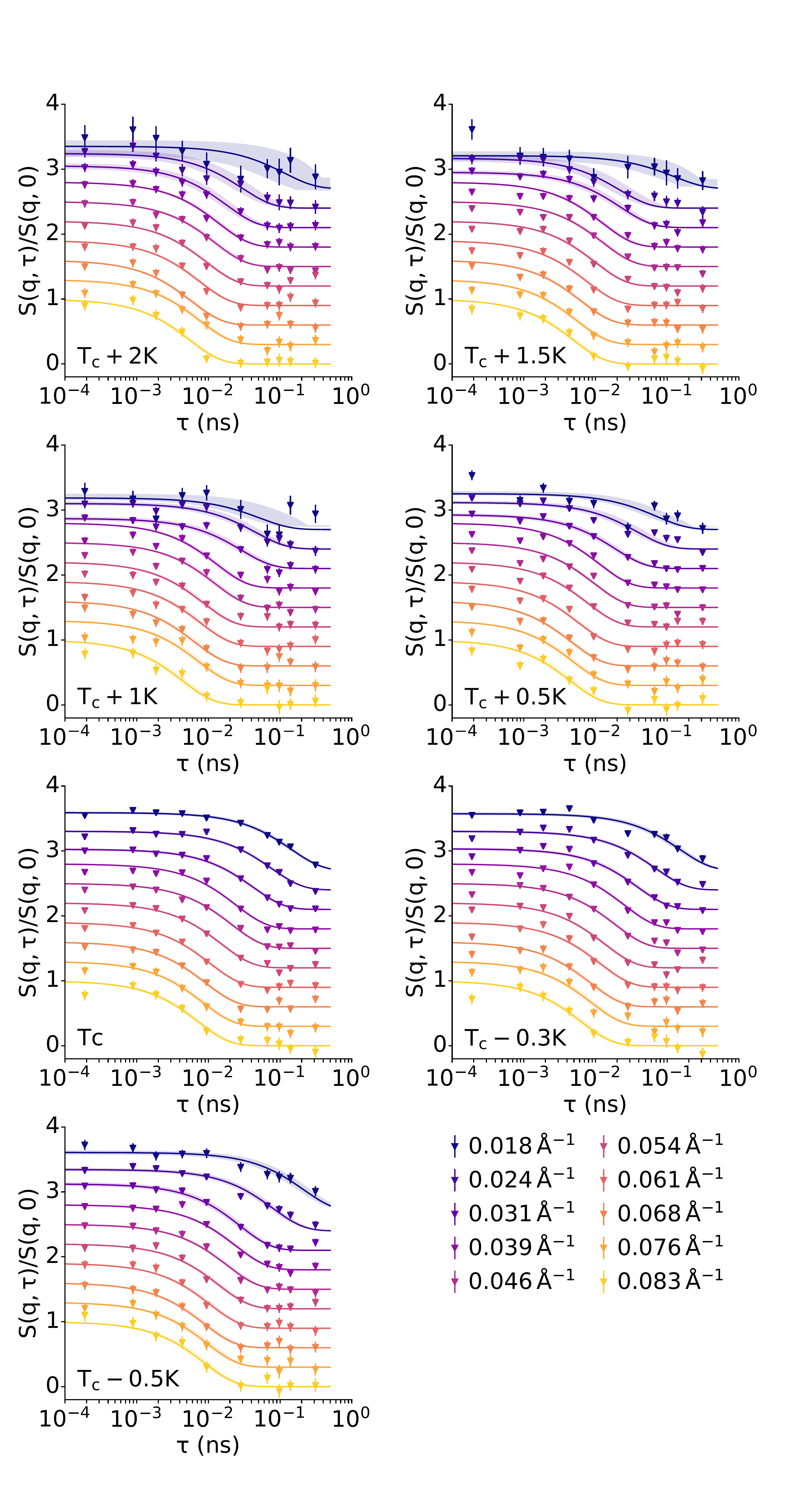}
	\caption{Intermediate scattering function $S(q,\tau)$ as a function of MIEZE time $\tau$ and for different temperatures T and wavevector transfers $q$ as determined for \ug. Solid lines are fits to Eq.\,\ref{eq:exp_fit} in the text. Shaded regions illustrate the uncertainty of the fit parameters.}
	\label{Fig:SpinEchoCurves}
\end{center}
\end{figure}

Here $\tau_{MIEZE}$ denotes the characteristic MIEZE time defined as
\begin{equation}
	\tau_{MIEZE} = \frac{m^2}{\pi h^2} L_{SD}(\omega_B - \omega_A) \lambda^3
\end{equation}
with the neutron mass $m$, the Planck constant $h$, the sample detector distance $L_{SD}$, the resonant flipper frequencies $\omega_{A, B}$ and the neutron wavelength $\lambda$. We show the obtained intermediate scattering function $S(q,\tau)$ for all measured momentum transfers $q$ and temperatures $T$ for \ug\ in Fig.\,\ref{Fig:SpinEchoCurves}. We note that $S(q,\tau)$ has been normalized to the static scattering at $S(q,0)$.

The scattering function for critical spin fluctuation has Lorentzian lineshape with line width $\Gamma_q$ [cf. Eq.~1 in the main text]
\begin{equation}
	S(q,\omega) \propto \Gamma_q/(\omega^2 + \Gamma_q^2)
\end{equation}
which corresponds to an exponential decay in the time-domain and we can fit the intermediate scattering function via 
\begin{equation}
	S(q,\tau)/S(q,0) = \exp(-|\Gamma_q| \cdot \tau_{MIEZE}).
	\label{eq:exp_fit}
\end{equation} 
The solid lines in Fig.\,\ref{Fig:SpinEchoCurves} denote the corresponding fits. The temperature and momentum transfer dependence of the linewidth $\Gamma_q(T)$ of the spin fluctuations determined via these fits is shown in Fig.~7(b) in the main text.

\section{Acknowledgments}

The authors wish to thank the technical staff at MLZ for their help in conducting the experiments. We are grateful to Olaf Soltwedel and Bj{\"o}rn Pedersen for assistance with the MIEZE and neutron Laue diffraction measurements, respectively. We also acknowledge useful discussions with J. M. Lawrence, F. Ronning, and P. B\"{o}ni. Work at Los Alamos National Laboratory (LANL) was supported by LANL Laboratory Directed Research and Development program. Work at Technische Universit\"{a}t M\"{u}nchen was supported by the TRR80 Project F2. The authors acknowledge the financial support by the Federal Ministry of Education and Research of Germany in the framework of 'Longitudinale Resonante Neutronen Spin-Echo Spektroskopie mit Extremer Energie-Auflösung' (project number 05K16WO6). The work by FH and MJ was supported through a Hans Fischer fellowship of the Technische Universit\"{a}t M\"{u}nchen --- Institute for Advanced Study, funded by the German Excellence Initiative and the European Union Seventh Framework Programme under grant agreement n$^{\circ}$ 291763. We also acknowledge support from the European Union through the Marie-Curie COFUND program.

\end{bibunit}


\begin{thebibliography}{10}

    \bibitem{Han:2012}
    T.-H. Han, J. S. Helton, S. Chu, D. G. Nocera, J. A. Rodriguez-Rivera, C. Broholm, and Y. S. Lee,Nature {\bf 492}, 406–410 (2012).

\bibitem{Fernandes:2014}
    R. M. Fernandes, A. V. Chubukov, and J. Schmalian,  
    Nature Physics {\bf 10}, 97–104 (2014).  

\bibitem{Muehlbauer:2009}
    S. Mühlbauer, B. Binz, F. Jonietz, C. Pfleiderer, A. Rosch, A. Neubauer, R. Georgii, P. Böni,
    Science {\bf 323}, 915-919 (2009).

\bibitem{Janoschek:2013}
    M. Janoschek, M. Garst, A. Bauer, P. Krautscheid, R. Georgii, P. Böni, and C. Pfleiderer,
    Phys. Rev B {\bf 87}, 134407 (2013).

\bibitem{Aaronson:1995}
    M. C. Aronson, R. Osborn, R. A. Robinson, J. W. Lynn, R. Chau, C. L. Seaman, and M. B. Maple,
    Phys. Rev. Lett. {\bf 75}, 725 (1995).

\bibitem{Schroeder:2000}
    A. Schröder, G. Aeppli, R. Coldea, M. Adams, O. Stockert, H.v. Löhneysen, E. Bucher, R. Ramazashvili, and P. Coleman,
    Nature {\bf 407}, 351–355 (2000). 

\bibitem{Kadowaki:2006}
    H. Kadowaki, Y. Tabata, M. Sato, N. Aso, S. Raymond, and S. Kawarazaki,
    Phys. Rev. Lett. {\bf 96}, 016401 (2006).

\bibitem{Knafo:2009}
    W. Knafo, S. Raymond, P. Lejay and J. Flouquet,
    Nature Physics {\bf 5} 753–757 (2009).
    
\bibitem{Scalapino:2012}
    D. J. Scalapino,
    Rev. Mod. Phys. {\bf 84}, 1383 (2012).        
    
\bibitem{Tsvelik:2003}
    A. M. Tsvelik,  Quantum Field Theory in Condensed Matter Physics, Cambridge University Press, (2003).

\bibitem{Dagotto:2005}
    E. Dagotto, 
    Science {\bf 309}, 257 (2005).       

\bibitem{Pfleiderer:2009}
	C. Pfleiderer, 
    Rev. Mod. Phys. {\bf 81}, 1551 (2009).
    
\bibitem{Moore:2009}    
    K. T. Moore and G. van der Laan,
    Rev. Mod. Phys. {\bf 81} 235 (2009).

\bibitem{Im:2008}
    H. J. Im, T. Ito, H.-D. Kim, S. Kimura, K. E. Lee, J. B. Hong, Y. S. Kwon, A. Yasui, and H. Yamagami,
    Phys. Rev. Lett. {\bf 100}, 176402 (2008)
    
\bibitem{Schmidt:2010}
    A. R. Schmidt,  M. H. Hamidian, P. Wahl, F. Meier, A. V. Balatsky, J. D. Garrett, T. J. Williams, G. M. Luke, and J. C. Davis,
    Nature {\bf 465}, 570 (2010).
   
\bibitem{Butch:2015}
    N. P. Butch, M. E. Manley, J. R. Jeffries, M. Janoschek, K. Huang, M. B. Maple, A. H. Said, B. M. Leu, and J. W. Lynn,
    Phys. Rev. B {\bf 91}, 035128 (2015).
    
\bibitem{Janoschek:2015}
    M. Janoschek, Pinaki Das, B. Chakrabarti, D. L. Abernathy, M. D. Lumsden, J. M. Lawrence, J. D. Thompson, G. H. Lander, J. N. Mitchell, S. Richmond, M. Ramos, F. Trouw, J.-X. Zhu, K. Haule, G. Kotliar, E. D. Bauer,
    Science Advances {\bf 1}, e1500188 (2015).
    
\bibitem{Goremychkin:2018}
    E. A. Goremychkin, H. Park, R. Osborn, S. Rosenkranz, J.-P. Castellan, V. R. Fanelli, A. D. Christianson, M. B. Stone, E. D. Bauer, K. J. McClellan, D. D. Byler, J. M. Lawrence,
	Science {\bf 359}, 186-191 (2018).
	
\bibitem{Gaehler:1992}	
	R. G{\"a}hler, R. Golub, and T. Keller, 
	Physica B {\bf 180}, 899 (1992).	

\bibitem{Saxena:2001}
	S. S. Saxena, P. Agrwal, A. Ahilan, F. M. Grosche, R. K. W. Haselwimmer, M. J. Steiner, E. Pugh, I. R. Walker, S. R. Julian, P. Monthoux, G. G. Lonzarich, A. Huxley, I. Sheiken, D. Braithwaite, and J. Flouquet,
    Nature {\bf 406}, 587 (2000).
    
\bibitem{Pfleiderer:2002}
	C. Pfleiderer and A. D. Huxley,
	Phys. Rev. Lett. {\bf 89} 147005 (2002).
	    
\bibitem{Fay:1980}
	D. Fay and J. Appel,
    Phys. Rev. B {\bf 22}, 3173 (1980).

\bibitem{Huxley:2003}
	A. D. Huxley, S. Raymond, and E. Ressouche,
	Phys. Rev. Lett. {\bf 91} 207201 (2003).	
    
\bibitem{Monthoux:1999}
    P. Monthoux and G. G. Lonzarich,
    Phys. Rev. B {\bf 59} 14598 (1999).   

\bibitem{HAUSSLER:2007}
	W. H{\"a}u{\ss}ler, B. Gohla-Neudecker, R. Schwikowski, D. Streibl and P. B{\"o}ni,
	Physica B {\bf 397}, 112 (2007).
	
\bibitem{Kindervater:2015}
	J. Kindervater, N. Martin, W. H{\"a}u{\ss}ler, M. Krautloher, C. Fuchs, S. M{\"u}hlbauer, J.A. Lim, E. Blackburn, P. B{\"o}ni, and C. Pfleiderer,
	EPJ Web of Conferences {\bf 83}, 03008 (2015).
	
\bibitem{Krautloher:2016}	
	Ma. Krautloher, J. Kindervater, T. Keller, and W. H{\"a}u{\ss}ler,
	Rev. Sci. Instr. {\bf 87}, 125110 (2016).

\bibitem{supplement}
	See Supplemental Material (attached) for the characterization of the sample with neutron Laue diffraction and neutron depolarization analysis, as well as detailed information about the used MIEZE setup, and additional SANS measurements. Further, we describe approximations used for the analysis of energy-integrated magnetic critical scattering and resolution calculations. Finally, we explain how the MIEZE data was analyzed. 	

\bibitem{Huy:2009}
	N. T. Huy {\it et al.}, 
	J. Magn. Magn. Mater. {\bf 321}, 2691 (2009).

\bibitem{Lynn:1998}
	J.W. Lynn, L. Vasiliu-Doloc, and M.A. Subramanian,
		Phys. Rev. Lett. {\bf 80}, 4582 (1998).

\bibitem{Simon:2002}	
	Ch. Simon, S. Mercone, N. Guiblin, C. Martin, A. Brulet, and G. Andre,
	Phys. Rev. Lett. {\bf 89}, 207202 (2002).
	
\bibitem{Chaikin}
	P. M. Chaikin and T. C. Lubensky,
	Principles of Condensed Matter Physics, Cambridge University Press (1995).	
    
\bibitem{Kernavanois:2005}
	N. Kernavanois, B. Grenier, A. Huxley, E. Ressouche, J. P. Sanchez, and J. Flouquet
	Phys. Rev. B 64, 174509 (2001).	    

\bibitem{Marshall:1968}
	W. Marshall and R. D. Lowde,
	Rep. Prog. Phys. {\bf 31} 705 (1968).
	
\bibitem{Squires:1978}
	G. L. Squires,
	Introduction to the theory of thermal neutron scattering, Dover Publications (1978).

\bibitem{Hohenberg:1977}
	P. C. Hohenberg and B. I. Halperin,
    Rev. Mod Phys., {\bf 49} 435 (1977).
    
\bibitem{Kindervater:2017}
	J. Kindervater, S. S{\"a}ubert, and P. B{\"o}ni,
    Phys. Rev. B {\bf 95}, 014429 (2017). 

\bibitem{Mirkiewicz:1969}
	V. J. Minkiewicz, M. F. Collins, R. Nathans, and G. Shirane
	Phys, Rev. {\bf 182}, 624 (1969).

\bibitem{Glitzka:1977}
	C. J. Glizka, V. J. Minkiewicz, and L. Passell,
	Phys. Rev. B {\bf 16}, 4084 (1977).

\bibitem{Dietrich:1976}
	O. W. Dietrich, J. Als-Nielsen, and L. Passell,
    Phys. Rev. B {\bf 14}, 4923 (1976).

\bibitem{Halperin:1969}
    B. I. Halperin and P. C. Hohenberg,
    Phys. Rev. 177, 952 (1969).    
	
\bibitem{Lonzarich:1986}
	G.G. Lonzarich,
	J. Magn. Magn. Mater. {\bf 54–57}, 612 (1986).

\bibitem{Bernhoeft:1988}	
	N. Bernhoeft, S. A. Law, G. G. Lonzarich, and D. McK. Paul,
	Phys. Scr. 38, 191 (1988).	

\bibitem{Lonzarich:1985}
	G.G. Lonzarich, L.Taillefer,
	J. Phys. C {\bf 18}, 4339 (1985).

\bibitem{Lonzarich:1999}
	G. G. Lonzarich,
	Electron: A Centenary Volume, edited by M. Springford, Cambridge University Press (1999), p. 109.

    \bibitem{Lashley:2006}
	J.C.Lashley, R.A.Fisher, J.Flouquet, F.Hardy, A.Huxley, N.E. Phillips,
	Physica B {\bf 378–380} 961-962 (2006).

\bibitem{Troc:2012}
	R. Tro\'{c}, Z. Gajek, and A. Pikul,
	Phys. Rev. B {\bf 86}, 224403 (2012).

\bibitem{Oikawa:1996}
	K. Oikawa, T. Kamiyama, H. Asano, Y. \={O}nuki, and M Kohgi,
	J. Phys. Soc. Jpn. {\bf 65} 3229 (1996).

\bibitem{Hill:1970}
	H. H. Hill,
	Plutonium and other Actinides, edited by W. N. Miner, The Metallurgical Society of AIME, New York (1970).

\bibitem{Tateiwa:2004}
	N. Tateiwa, T. C. Kobayashi, K. Amaya, Y. Haga, R. Settai, and Y. \~{O}nuki,
	Phys Rev B {\bf 69} 180513 (2004).

\bibitem{Terashima:2001}
	T. Terashima, T. Matsumoto, C. Terakura, S. Uji, N. Kimura, M. Endo, T. Komatsubara, and H. Aoki,
	Phys. Rev. Lett. {\bf 87}, 166401 (2001).

\bibitem{Settai:2002}
	R. Settai M. Nakashima, S. Araki, Y. Haga., T. C. Kobayashi, N. Tateiwa, H. Yamagami, and Y. Onuki,
	J. Phys.: Condens. Matter {\bf 14} L29–L36 (2002).
	
\bibitem{Thomas:2011}
	C. Thomas, A. S. da Rosa Sim\~{o}es, J. R. Iglesias, C. Lacroix, N. B. Perkins, and B. Coqblin, 
	Phys. Rev. B, {\bf 83} 014415 (2011).

\bibitem{Hoshino:2013}
	S. Hoshino and Y. Kuramoto, 
	Phys. Rev. Lett., {\bf 111} 026401, (2013).

\bibitem{Hattori:2012}
	T. Hattori, Y. Ihara, Y. Nakai, K. Ishida, Y. Tada, S. Fujimoto, N.
	Kawakami, E. Osaki, K. Deguchi, N. K. Sato, and I. Satoh,
	Phys. Rev. Lett. 108, 066403 (2012).
	
\bibitem{Stock:2011}
	C. Stock, D. A. Sokolov, P. Bourges, P. H. Tobash, K. Gofryk, F. Ronning, E. D. Bauer, K. C. Rule, and A. D. Huxley,
	Phys. Rev. Lett. 107, 187202 (2011).

\bibitem{Kepa:2014}
	M. W. Kepa, D. A. Sokolov, M B\"{o}hm and A. D. Huxley,
	J. Phys.: Conf. Ser. {\bf 568} 042016 (2014).

\bibitem{Martin:2018}
	N. Martin,
	Nuclear Inst. and Methods in Physics Research, A {\bf 882} 11–16 (2018).  


\end{thebibliography}

\begin{thebibliography}{10}
\bibitem{Schulz:2010}
	M. Schulz, A. Neubauer, S. Masalovich, M. M{\"u}hlbauer, E. Calzada, B. Schillinger, C. Pfleiderer, P. and B{\"o}ni, 
	J. Phys.: Conf. Ser. {\bf 211} 012025 (2010).
	
\bibitem{Seifert:2017}
	M. Seifert, M. Schulz, G. Benka, C. Pfleiderer and S. Gilder,
	J. Phys.: Conf. Ser. {\bf 862}, 012024 (2017).
	
\bibitem{Muehlbauer:2016}
	S. M{\"u}hlbauer, {\it et al.}, 
	NIMA 832, 297-305 (2016).

\bibitem{Huxley:2003}
	A. D. Huxley, S. Raymond, and E. Ressouche,
	Phys. Rev. Lett. {\bf 91} 207201 (2003).

\bibitem{Marshall:1968}
	W. Marshall and R. D. Lowde,
	Rep. Prog. Phys. {\bf 31} 705 (1968).
	
\bibitem{Squires:1978}
	G. L. Squires,
	Introduction to the theory of thermal neutron scattering, Dover Publications (1978).
	
\bibitem{Pederson:1990}
	J. S. Pedersen and D. Posselt and K. Mortensen, 
	J. Appl. Cryst. {\bf 23}, 321 (1990).
	
\bibitem{2015_Franz_JLRFJ}
	C. Franz, and T. Schr{\"o}der,
	Journal of large-scale research facilities {\bf 1} 37 (2015).

\bibitem{2002_Keller_S}
	T. Keller, R. Golub, Robert and R. G{\"a}hler,
	Scattering, Academic Press (2002), p. 1265.

\bibitem{Klein:2009}	
	M. Klein and C. Schmidt, 
	Nucl. Instrum. Methods Phys. Res., Sect. A {\bf 628}, 9 (2011).

\bibitem{HAUSSLER:2011}
	W. H{\"a}u{\ss}ler, P. B{\"o}ni, M. Klein, C. Schmidt, U. Schmidt, F. Groitl, and J. Kindervater,
	Rev. Sci. Instrum. {\bf 82}, 045101 (2011).	

\bibitem{2005_Haussler_PCCP}
	W. H{\"a}ussler, and U. Schmidt,
	Phys. Chem. Chem. Phys. {\bf 7} 1245 (2005).
\end{thebibliography}
\end{document}